\documentclass[journal,draftcls, onecolumn,12pt,twoside]{IEEEtranTCOM}
\usepackage{amsmath}
\usepackage{amsfonts}
\usepackage{mathtools} % ceil,floor,etc
\usepackage{epsfig,graphicx}
\usepackage{color} % textcolor
\usepackage{epstopdf}
\usepackage{verbatim}  % multi-line comments
\usepackage[]{hyperref}
\usepackage{xcolor}
\hypersetup{  %  surten els colors als links pero no tan lletjos
    colorlinks,
    linkcolor={blue!60!black},
    citecolor={blue!60!black},
    urlcolor={blue!80!black}
}

\usepackage{pmat} % matrius amb ratlles per separar blocs
\usepackage{booktabs} % taules
\usepackage{hhline} % taules
\usepackage{tikz}  % per carregar imatges de matlab fetes latex
\usepackage{pgfplots}  
\usetikzlibrary{plotmarks}
\usepackage{nicefrac}
\usepackage{setspace}

\usepackage[nohyperlinks]{acronym}
\newcommand{\acronim}[2]{  \newacro{#1}[#1]{#2}
}

\newcommand{\pare}[1] {\left( #1 \right)} % parentesis
\newcommand{\parep}[1] {\big( #1 \big)} % parentesis
\newcommand{\etal}{\textit{et al.}\,\,} 

\newcommand{\Cmat}[2]{\!\in\mathbb{C}^{#1 \times #2}}
\newcommand{\Rmat}[2]{\!\in\!\mathbb{R}^{#1 \times #2}}
\newcommand{\Span}[1]{\mathsf{span} \big( #1 \big)}
\newcommand{\Rank}[1]{\mathsf{rank} \pare{#1}}
\newcommand{\Null}[1]{\mathsf{null} \big( #1 \big)}
\newcommand{\Stack}[1]{\mathsf{stack} \Big(  #1  \Big)}
\newcommand{\Text}[1]{\mathsf{T} \pare{#1}}
\newcommand{\ACS}[1]{\mathsf{ACS} \pare{#1}}
\newcommand{\CB}[1]{\mathsf{CB} \parep{#1}}
\newcommand{\C}[1]{\Cn \pare{#1}} % matriu C(a)
\newcommand{\hel}[2]{ {h}_{#1}^{#2} } % matriu C(a)
\newcommand{\Ch}[2]{ \C{ \hel{#1}{#2} } } % matriu C(a)
\newcommand{\Gel}[2]{\C{ \hat{\theta}_j^{#1}} \An_{#2} }
\newcommand{\Geldes}[2]{ \Gdes(#1) \An_{#2} }

\newcommand{\elim}[2]{\psi \left( #1,#2 \right)}

\DeclareMathOperator{\Vn}{\mathbf{V}}
\DeclareMathOperator{\Hn}{\mathbf{H}}
\DeclareMathOperator{\Gn}{\mathbf{G}}
\DeclareMathOperator{\Gdes}{\Gn_\mathit{j}^{\text{des}}}
\DeclareMathOperator{\Gint}{\Gn_\mathit{j}^{\text{int}}}
\DeclareMathOperator{\Cn}{\mathbf{C}}
\DeclareMathOperator{\An}{\mathbf{A}}
\DeclareMathOperator{\0}{\mathbf{0}}
\DeclareMathOperator{\I}{\mathbf{I}}
\DeclareMathOperator{\G0}{\mathbf{0}}

\newtheorem{theorem}{Theorem}
\newtheorem{lemma}{Lemma}

\newtheorem{conjecture}{Conjecture}

\IEEEoverridecommandlockouts

\renewcommand{\IEEEQED}{\IEEEQEDopen} % signe de proof blanc
\begin{document}

\copyright 2014 IEEE. Personal use of this material is permitted. Permission from IEEE must be 
obtained for all other uses, in any current or future media, including 
reprinting/republishing this material for advertising or promotional purposes, creating new 
collective works, for resale or redistribution to servers or lists, or reuse of any copyrighted 
component of this work in other works. 

\newpage

%\begin{acronym}
	\acronim{ACS}{Asymmetric Complex Signaling}
	\acronim{BC}{Broadcast channel}
	\acronim{BS}{Base station}
	\acronim{CSI}{Channel State Information}
	\acronim{CSIT}{Channel State Information at the transmitter side}
	\acronim{DoF}{Degrees of Freedom}
	\acronim{i.i.d.}{independent and identically distributed}
	\acronim{IA}{Interference Alignment}
	\acronim{IC}{Interference Channel}
	\acronim{MAC}{Multiple access channel}
	\acronim{MIMO}{Multiple-Input Multiple-Output}
	\acronim{pdf}{probability density function}
	\acronim{SISO}{Single-Input Single-Output}
	\acronim{SNR}{Signal to Noise Ratio}
	\acronim{UE}{User equipment}
	\acronim{ZF}{Zero forcing}
	\acronim{ZP}{Zero Propagation}
%\end{acronym}

% paper title
% can use linebreaks \\ within to get better formatting as desired
\title{The DoF of the 3-user ($p,p+1$) MIMO Interference Channel}
% author names and IEEE memberships
% note positions of commas and nonbreaking spaces ( ~ ) LaTeX will not break
% a structure at a ~ so this keeps an author's name from being broken across
% two lines.
% use \thanks to gain access to the first footnote area
% a separate \thanks must be used for each paragraph as LaTeX2e's \thanks
% was not built to handle multiple paragraphs

\author{Marc Torrellas, Adrian Agustin, Josep Vidal and Olga Mu\~noz
\thanks{Manuscript submitted on June 9, 2014 to the IEEE Transactions on Communications. This work has been done in the framework of the projects TROPIC FP7 ICT-2011-8-318784, funded by the European Commission, MOSAIC (TEC2010-19171/TCM) and DISNET (TEC2013-41315-R), funded by Spanish Ministry of Education, and  2014SGR-60, funded by the Catalan Administration.}}

% The paper headers
\markboth{IEEE Transactions on Communications}%
{Submitted paper}

\maketitle

\begin{abstract}
The \emph{degrees of freedom} (DoF) of the \mbox{3-user} multiple-input multiple-output (MIMO) interference channel (IC) with full channel state information (CSI) and constant channel coefficients are characterized when linear filters are employed and $(p,p+1)$ antennas are deployed at the transmitters and receivers, respectively. The point of departure of this paper is the work of Wang \etal \cite{AlignmentChainsTrans}, which conjectured but not proved the DoF for the configuration tackled in this work. In this work we prove the optimal DoF by means of a transmission scheme based on asymmetric complex signaling (ACS) together with symbol extensions in time and interference alignment (IA) concepts. The paper deals with the $p=2,3$ cases, providing the transmit and receive filters and the tools needed for proving the achievability of the DoF for other values of $p$.
\end{abstract}

% Note that keywords are not normally used for peerreview papers.

\IEEEpeerreviewmaketitle

\section{Introduction}
 
In recent years, the {\it degrees of freedom} (DoF) have emerged as one of the most important metrics for characterizing interference networks. The DoF describe how the system sum rate scales with the logarithm of the signal to noise ratio (SNR) at the high SNR regime. The Interference Alignment (IA) concept has elucidated the optimal DoF for certain configurations of interference networks. The main purpose of \ac{IA} is to design the transmit filters in such a way that each receiver observes all the interfering signals overlapped in a common subspace. 
The concept was originally proposed in the context of index coding in \cite{Birk&Kol},
while it crystallized later on for the 2-user \ac{MIMO} X-channel in \cite{Maddah-Ali2008} and for the $K$-user \ac{SISO} \ac{IC} with $K>2$ in \cite{CJ}. Surprisingly, Cadambe \etal \cite{CJ} proposed a linear precoding scheme that provides \textit{each user half the cake}, and therefore a total of $\nicefrac{K}{2}$ DoF over the network. Additionally, the authors showed that this result generalizes to the \ac{MIMO} case, obtaining $\nicefrac{KM}{2}$ total DoF when all nodes are equipped with $M$ antennas. For both cases, the achievability of fractional DoF relies on transmitting along an arbitrarily large number of channel uses on a time-varying / frequency-selective channel. However, it fails when considering a constant \ac{SISO} channel because the equivalent channel matrices result on scaled identity matrices and the diversity provided by the channel variations cannot be exploited. 

There is a large number of works in the literature that have employed the IA concept for analyzing the MIMO \ac{IC} in terms of DoF, see for example \cite{AlignmentChainsTrans,Yetis2010,Razaviyayn2012,Gou2010_DoF_MN,GenieChainsArxiv,Reciprocity,ACS,Geometry2011}. Especially interesting is the work in \cite{Reciprocity}, where the authors showed the DoF reciprocity concept in wireless networks \cite{Reciprocity}. This property states that given a network with one particular antenna setting, its reciprocal setting, i.e. a network where the number of antennas at the transmitters and receivers (or the roles of transmitters and receivers) are exchanged, has exactly the same DoF. This important result allows to half the number of antenna settings to be investigated.

After the disrupting idea of IA, different types of IA emerged. Basically, there are two different frameworks for developing IA-based transmit precoders: lattice level IA \cite{LayeredIA} (\emph{lattice alignment}), and vector space level IA \cite{Maddah-Ali2008}, (\emph{vector space alignment}). These two techniques arise from the choice between structured or random codes, respectively. Lattice alignment-based techniques exploit the rational dimensions framework such that the undesired signals are seen at each receiver on the same lattice. Inspired on this idea, \cite{Ghasemi2010} showed that the DoF outer bound may be attained for almost any user and antenna settings. Nevertheless, this type of IA is highly dependent on the SNR and its rate performance is extremely degraded at medium SNR \cite{Ordentlich2013}.

On the other hand, vector space alignment techniques are able to attain the optimal DoF only for certain antenna configurations, but in contrast to the lattice-based techniques, they show a better rate performance at medium SNR. Basically, there exist in the literature 3 types of IA in the vector space alignment framework currently proposed: the conventional approach explained before, Ergodic IA (EIA)\cite{ErgodicIA} and Opportunistic IA (OIA) \cite{OIA}. The idea of EIA relies on repeating the same transmission along two time slots with complementary channel states, such that by summing up the signals received from both time slots the interference is canceled. The surprising result in \cite{ErgodicIA} was that each user attains the $\frac{1}{2}$ DoF in a $K$-user SISO IC regardless the number of users $K$, without the need of any precoding at the transmitter side, and for any SNR value. However, it was also shown that the delay users must wait for complementary channel states grows as the SNR. For this reason, we will assume that transmitters cannot choose the time slots when they transmit, and ergodic IA will not be considered. On the other hand, OIA exploits the user dimension through scheduling. The idea is to combine the benefits of opportunistic beamforming and IA. The advantage of this approach is that information sharing among transmitters is not required, and the required CSI feedback may be highly alleviated. However, the number of users associated to each transmitter should grow with the SNR \cite{OIA}.

This paper will consider the conventional IA approach. In this context, the best known inner bound for the 3-user SISO IC was proposed by Cadambe \etal in \cite{ACS}. The authors proposed a linear precoding scheme able to achieve $1.2$ DoF, thanks to the \textit{asymmetric complex signaling} (ACS) concept. This approach, together with symbol extensions in time, is able to exploit the real and imaginary components of the channel. As a result, the equivalent channel matrices are no longer scaled identity matrices but present a more sophisticated structure that can be exploited by the IA scheme. A similar approach has been recently reported for the 4-user SISO IC in \cite{ACS_4users}. 

Moreover, Wang \etal have recently characterized the 3-user MIMO IC \cite{AlignmentChainsTrans} in DoF terms. On the one hand, the DoF outer bound is derived by introducing the change of basis (CB) operation, which allows to write the equivalent channels in such a way that the appropriate genie signals to be provided to each receiver can be more easily identified\footnote{Interestingly, the CB operation has been found to be useful for other settings, e.g. the MIMO rank-deficient IC \cite{Zeng_defRank_MIMO3}.}. On the other hand, the proposed DoF inner bound flows from the \textit{subspace alignment chains} concept. This approach proposes a transmitter design intertwined among users through the alignment, being optimal in DoF terms for almost all antenna settings. Nevertheless, the SISO case and all $\left(p+1, p \right)$ and $\left(p, p+1 \right)$ MIMO cases with $p\!>\!1$\footnote{The case $p=1$ was previously addressed in \cite{Gou2010_DoF_MN}.} remain open problems (see Section 8.3 in \cite{AlignmentChainsTrans}) when the channel coefficients are held constant. In this regard, it is worth pointing out that the DoF characterization of the 3-user $\left(p, p+1 \right)$ MIMO IC has been later on claimed in \cite{AlignmentChains_ISIT,ACSnoPublicat} by means of \ac{ACS} and \textit{subspace alignment chain} concepts, but the result is just sustained on numerical experiments. Therefore, to the best of the authors' knowledge, there is not a formal proof in the literature. 

\begin{figure}[]
%\centering
\centerline{\includegraphics[width=0.65\linewidth]{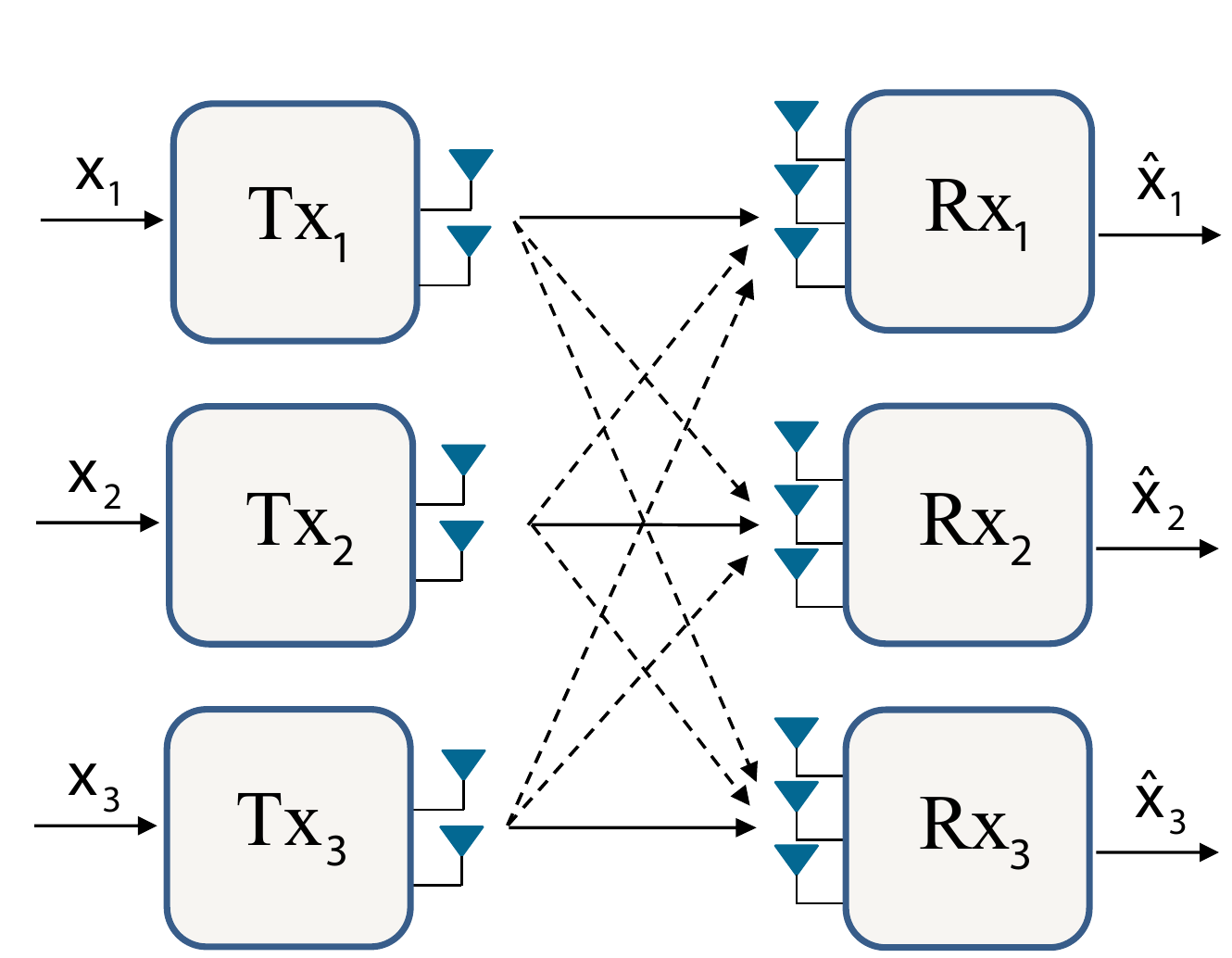}}
\caption[The 3-user MIMO IC]{The 3-user (2,3) MIMO IC. Transmitters and receivers are equipped with $p=2$ and $p+1=3$ antennas, respectively. Solid lines define the intended signals, while dotted lines denote the interfering signals.}
\label{fig:IC}
\end{figure}  

\subsection{Contributions}
\label{sec:contributions}
The goal of this work is to provide a formal proof about the optimal DoF of the 3-user $\left(p,p+1 \right)$ MIMO IC with constant channel coefficients using linear transmit-receive filters. As an example of such scenarios, Fig.\ref{fig:IC} shows the $\text{3-user} \left(2, 3 \right)$ MIMO IC. The proposed scheme is based on interference alignment, \textit{symbol extensions in time} and \textit{asymmetric complex signaling}. Three contributions summarize this work:

\begin{itemize}
 \item We prove that the DoF per user for ${p=2,3...6}$ are exactly $\frac{p\left(p+1\right)}{2p+1}$, see Theorem \ref{theorem23} and \mbox{Theorem \ref{theorem34}} in Sections \ref{sec:5case23} and \ref{sec:6generalCase}, respectively.
 \item The proposed transmit precoding matrices present a specific structure that can be generalized for any value of $p$. This structure is characterized by two properties: {\it i)} there are some elements equal to zero, and {\it ii)} all transmit precoders are defined as a function of 3 matrices, denoted as the \textit{support precoding blocks}. An iterative algorithm is proposed, able to find the structure of each precoding matrix for any value of $p$. 
 \item By generalizing the proposed scheme and proof methodology, we conjecture that the 3-user $\left(p,p+1 \right)$ MIMO IC with constant channel coefficients has $\frac{p\left(p+1\right)}{2p+1}$ DoF per user for any ${p\geq7}$, achieved by means of linear filters at transmitters and receivers.
\end{itemize}

\subsection{Organization}
This paper is organized as follows. \mbox{Section \ref{sec:2SystemModel}} introduces the system model considered in this work. Next, Section \ref{sec:DoF} reviews the DoF for our specific scenario, as well as DoF achievability conditions when using IA. The structure of our precoding scheme and the alignment chains are addressed in Section \ref{sec:PrecodingStructure}. Section \ref{sec:5case23} is devoted to present the $p=2$ case, while Section \ref{sec:6generalCase} addresses the $p=3$ case, which differs from the previous case in notation, and allows the generalization of the precoding scheme for $p>3$. This is achieved by means of the $\textit{zero propagation algorithm}$, presented in \mbox{Section \ref{sec:6generalCase}}. These cases $p=2,3$ allow understanding the achievability proof for the general case. Moreover, simulation results are provided in \mbox{Section \ref{sec:4.3sim}}, where the the sum-rate is depicted as a function of the SNR for different values of $p$, and DoF achievability is shown. Finally, conclusions are drawn in \mbox{Section \ref{sec:7conclusion}}.

\subsection{Notation}
We write vectors in boldface lowercase types ($\mathbf{x}$), and matrices in boldface uppercase types ($\mathbf{X}$). $(\cdot)^T$,  $(\cdot)^H$, and $\otimes$ stand for the transpose, transpose and conjugate, and Kronecker product operators, respectively, and we define
\begin{IEEEeqnarray}{c}
\Stack {
  \mathbf{A} , \mathbf{B}  } = 
\Big[ 
 {\mathbf{A}}^T \quad {\mathbf{B}}^T 
\Big]^T.
\end{IEEEeqnarray}

Furthermore, for any given $N$-column vector ${\mathbf{x}=\left[x(1), x(2), \ldots, x(N)\right]^T}$ and ${M\text{-column}}$ matrix ${\mathbf{Y}=\left[ \mathbf{y}_1, \mathbf{y}_2, \ldots , \mathbf{y}_M \right]}$, we define 
\begin{IEEEeqnarray*}{c}
\mathbf{x}(a:b)=\left[x(a), x(a+1),\ldots,x(b)\right]^T, \quad
\mathbf{Y}_{a:b}=\left[\mathbf{y}_{a}, \mathbf{y}_{a+1}, \ldots, \mathbf{y}_{b}\right].
\end{IEEEeqnarray*}
Additionally, $\lceil . \rceil$, $\lfloor . \rfloor$, and $\langle . \rangle$ stand for the ceiling, floor, and modulo-3 
operators, respectively. We remark that all indices in this work are assumed to be in the set $\{1,2,3\}$, applying the modulo-3 operation only if necessary. Furthermore, $\Span{\mathbf{A}}$ defines the subspace generated by all linear combinations of the columns of $\mathbf{A}$, and $\Rank{\mathbf{A}}$ denotes its dimension. Finally, $ \mathbb{R},\mathbb{C}$ stand for the real and complex sets of numbers, respectively.

%%%%%%%%%%%%%%%%%%%%%%%%%%%%%%%%%%%%%%%%%%%%%%%%%%%%%%%%%%%%%%%%%%%%%%%%%%%%%%%%%%%%%%%%%%%%%%%%%%%%%%%%%%%%%%%%%%%%%%%%%%%%%%%%%%%%%%%%%%%%%%%%%%%%%%%%%%%%%%%%%%%%%%%%%%%
%%%%%%%%%%%%%%%%%%%%%%%%%%%%%%%%%%%%%%%%%%%%%%%%%%%%%%%%%%%%%%%%%%%%%%%%%%%%%%%%%%%%%%%%%%%%%%%%%%%%%%%%%%%%%%%%%%%%%%%%%%%%%%%%%%%%%%%%%%%%%%%%%%%%%%%%%%%%%%%%%%%%%%%%%%%
%%%%%%%%%%%%%%%%%%%%%%%%%%%%%%%%%%%%%%%%%%%%%%%%%%%%%%%%%%%%%%%%%%%%%%%%%%%%%%%%%%%%%%%%%%%%%%%%%%%%%%%%%%%%%%%%%%%%%%%%%%%%%%%%%%%%%%%%%%%%%%%%%%%%%%%%%%%%%%%%%%%%%%%%%%%

\section{System Model}
\label{sec:2SystemModel}
The 3-user $\left(p,p+1 \right)$ MIMO IC is considered, where each transmitter and each receiver is equipped with $p$ and $p+1$ antennas, respectively. Each transmitter aims to deliver a message to one unique receiver, labelled with the same index. Perfect and instantaneous \ac{CSI} is assumed and exploited at both sides. Channel coefficients are randomly drawn from some continuous complex probability density function, and assumed to be constant along the whole transmission time. The transmission is carried out over $2T$ equivalent channel uses thanks to the $T$ \textit{symbol extensions in time} and \textit{asymmetric complex signaling} \cite{ACS}. The received and processed signals may be written as follows
\begin{IEEEeqnarray}{c}
\label{eq:extendedSignal2}
\mathbf{{y}}_j = 
{{\mathbf{H}}_{j,j}} {{\mathbf{V}}_j}{{\mathbf{x}}_j} + \!\!
\sum\limits_{i = 1, i \ne j}^{3} {\mathbf{ H}}_{j,i} {\mathbf{V}}_i {\mathbf{x}}_i + {{\mathbf{n}}_j} 
\\
{{\mathbf{z}}_j} = {\mathbf{W}}_j {{\mathbf{y}}_j},
\label{eq:ProcessedSignal}
\end{IEEEeqnarray}
where ${\mathbf{y}}_j\Rmat{2T(p+1)}{1}$ is the received signal vector at the $j$th receiver, ${{\mathbf{z}}_j} \Rmat{\hat d_j}{1}$ is the processed signal, ${\mathbf{x}}_j \Rmat{\hat d_j}{1}$ is the vector with uncorrelated components composed of $\hat d_j$ real-valued data symbols defining the message intended to the $j$th receiver, ${\mathbf{V}}_j\Rmat{2Tp}{\hat d_j}$ is the precoding matrix of the $j$th transmitter, ${\mathbf{W}}_j \Rmat{\hat d_j}{2T(p+1)}$ is a linear receiving filter, and ${\mathbf{n}}_j\Rmat{2T(p+1)}{1}$ denotes the noise vector at the $j$th receiver, whose components are i.i.d. as $\mathcal{N}\!(0,1)$. Furthermore, $\mathbf{H}_{j,i} \Rmat{2T(p+1)}{2Tp}$ stands for the equivalent channel matrix from the $i$th transmitter to the $j$th receiver after considering symbol extensions in time and ACS concepts, and applying a CB operation, to be detailed next.

Let $\bar{\mathbf{H}}_{j,i} \Cmat{p+1}{p}$ be the original channel matrix from the $i$th transmitter to the $j$th receiver, and assume a transmission over $T$ channel uses. In such a case, one could stack all the received signals, and write a more compact system model. Therefore, the equivalent channel matrix could be written as follows:
\begin{IEEEeqnarray}{c}
\label{eq:TimeExtension}
\Text{ \mathbf{\bar{H}}_{j,i} } =  \mathbf{I}_T \otimes \bar{\mathbf{H}}_{j,i},
\end{IEEEeqnarray}
where $\mathbf{I}_T \, \Rmat{T}{T}$ is the identity matrix. 

Similarly, real and imaginary parts of the received signals could be considered separately, as done in \cite{ACS}. However, in contrast to \cite{ACS}, here we use the ACS concept for each particular channel coefficient. The extended form for each channel element is therefore written as:
\begin{IEEEeqnarray}{c}
\label{eq:ACSextension}
  \ACS{\bar{h}_{j,i}^{q,r}} =\left| \bar{h}_{j,i}^{q,r} \right| \bar{\mathbf{U}}\left( \bar{\phi}_{j,i}^{q,r} \right) 
\Rmat{2}{2},
\end{IEEEeqnarray}
where $\bar{h}_{j,i}^{q,r}$ is the \textit{complex} channel gain between the $r$th antenna of transmitter $i$ and the $q$th antenna of receiver $j$, $\bar \phi _{j,i}^{q,r}$ is the complex phase of 
$\bar{h}_{j,i}^{q,r}$, and 
$\mathbf{\bar U} \left( {\bar \phi _{j,i}^{q,r}} \right) \,\Rmat{2}{2} $ is an unitary matrix given by:
\begin{IEEEeqnarray}{c}
\label{eq:matrixU}
\bar{\mathbf{U}}\left( \bar \phi _{j,i}^{q,r} \right) = \begin{bmatrix}
	\cos \left( \bar \phi _{j,i}^{q,r} \right) 
	& - \sin \left( {\bar{\phi}_{j,i}^{q,r}} \right)
	\vspace{1mm}\\
	\sin \left( \bar {\phi} _{j,i}^{q,r} \right)
	& \cos \left( {\bar \phi _{j,i}^{q,r}} \right)
\end{bmatrix},
\end{IEEEeqnarray}
with some interesting properties, for example:
\begin{IEEEeqnarray}{l}
\label{eq:propertiesU}
\hskip -5mm
\begin{matrix}
{\mathbf{\bar U}}\left( a \right){\mathbf{\bar U}}\left( b \right) = { \mathbf{\bar U}}\left( {a+b} \right) ,
\,\,
{\mathbf{\bar U}}{\left( a \right)^{ - 1}} = {\mathbf{\bar U}}\left( -a \right),
\end{matrix}
\end{IEEEeqnarray}
for any arbitrary phases $a,b \in \left[ \,0,2\pi \,\right] $. 

For the sake of clarity, let us write the equivalent channel channel matrix ${\hat \Hn_{j,i} \Rmat{2T(p+1)}{2Tp}}$ when the two previous concepts are together applied, given by

\begin{IEEEeqnarray}{c}
\label{eq:extendedChannel}
\hat \Hn_{j,i} = 
\begin{bmatrix}
	\Cn \left( \bar{h}_{j,i}^{1,1} \right) 
	&
	\dots
	&
	\Cn \left(  \bar{h}_{j,i}^{1,p} \right) 
	\\
	\vdots & \ddots & \vdots
	\\
	\Cn \left( \bar{h}_{j,i}^{p+1,1} \right) 
	&
	\dots
	&
	\Cn \left( \bar{h}_{j,i}^{p+1,p} \right)
\end{bmatrix},
\end{IEEEeqnarray}
with
${\Cn \left(  \bar{h}_{j,i}^{q,r} \right) =  \left| \bar h_{j,i}^{q,r} \right| \mathbf{I}_T \otimes \mathbf{\bar U} 
\left( \bar \phi _{j,i}^{q,r} \right)}$. Now the last step to obtain the system model in (\ref{eq:extendedSignal2}) consists on applying a CB operation \cite{AlignmentChainsTrans}:
\begin{IEEEeqnarray}{c}
\label{eq:CB}
\Hn_{j,i} = \CB{ \hat\Hn_{j,i} } = \mathbf{R}_{j} \hat\Hn_{j,i} \mathbf{T}_{i},
\end{IEEEeqnarray}
where $\mathbf{R}_{j} \Rmat{2T(p+1)}{2T(p+1)} $ and $\mathbf{T}_{j} \Rmat{2Tp}{2Tp} $ are invertible linear transformations applied at the transmitters and the receivers. This way the equivalent channel becomes a rotation of $\hat\Hn_{j,i}$ with zeros at some specific antenna elements, see \cite{AlignmentChainsTrans} for details.

In this work, the same CB as in \cite{AlignmentChainsTrans} is applied at the transmit side, whereas that for the receiver side contains some additional operations described in \mbox{Appendix \ref{sec:appendixA}}. This way we obtain a simplified structure for the channel matrices, which simplifies the precoding design based on interference alignment, as well as the achievability proof.
\\
\textit{Remark}: Notice that matrices $\mathbf{R}_{j}$ and $\mathbf{T}_{j}$ are applied at the transmitter and the receiver, respectively. Therefore, the equivalent precoding matrix at each transmitter and receiving filter at each receiver are $\mathbf{W}_{j} \mathbf{R}_{j}$ and $\mathbf{T}_{j} \Vn_{j}$, respectively.

\section{Degrees of Freedom}
\label{sec:DoF}
The DoF per user $d_{j}$ for the 3-user $\left(p,p+1 \right)$ MIMO IC are upper bounded \cite{AlignmentChainsTrans} by 
\begin{IEEEeqnarray}{c}
\label{eq:DoFouterbound}
{d}_j \leq \dot{d}_{j} = \frac{{p\left( {p + 1} \right)}}{{2p + 1}}.
\end{IEEEeqnarray} 

On the other hand, the DoF achieved by the $j$th user assuming the channel model described in Section \ref{sec:2SystemModel} are given by
\begin{IEEEeqnarray}{c}
\label{eq:DoF_extended}
\mathring{d}_j = \frac{1}{2T}
\Rank{ \mathbf{W}_j \Hn_{j,j} \Vn_j } \overset{(a)}{\leq}
\frac{\hat{d}_j}{2T} {\leq} d_{j}
\end{IEEEeqnarray}
in case all the received interference is completely removed, i.e.
\begin{IEEEeqnarray}{c}
\mathbf{{W}}_{j} \mathbf{{H}}_{j,i} \mathbf{{V}}_i  = \0 , \quad \forall i \neq j .
\label{eq:noInterferenceb}
\end{IEEEeqnarray}
The previous condition forces ${\mathbf{W}}_j$ to be an orthogonal projection onto the interference space.  Consequently, (\ref{eq:DoF_extended} - $a$) will be satisfied with equality only in case the desired and interfering signals are \textit{linearly independent}. Let define the $\textit{signal space matrix}$ (SSM) as the matrix whose columns generate the sum space of desired and interference subspaces at each receiver,
\begin{IEEEeqnarray}{c}
\label{eq:Gdefinition}
\begin{matrix}
\Gn_j =
\big[
 \Gdes \, \Gint
\big]
\\[2mm]
\Span{\Gdes} = \Span{{\Hn}_{j,j}{\Vn}_j} 
\\[2mm]
\Span{\Gint}= \Span{ 
	\begin{bmatrix}
		{\Hn}_{j,j-1}{\Vn}_{j-1} &
		{\Hn}_{j,j+1}{\Vn}_{j+1}
	\end{bmatrix} },
\end{matrix}
\end{IEEEeqnarray}
where $\Gdes$ and $\Gint$ are defined as some full-rank matrices whose columns form a basis (see definition of operator $\mathsf{span}(\cdot)$ in the notation section) for the subspaces occupied by desired and interference signals, respectively. Given this formulation, proving the DoF achievability reduces to prove that the SSM is full-rank, since in such a case desired and interfering signals are linearly independent, thus there exists a solution for transmitting and receiving filters simultaneously satisfying  (\ref{eq:DoF_extended} - $a$) with equality and (\ref{eq:noInterferenceb}). 

The present work proves that the achievable DoF $\mathring{d}_{j}$ and outer bound DoF $\dot{d}_{j}$ coincide on the optimal DoF $d_{j}$. This is shown by proposing a precoding scheme that can reliably transmit ${\hat d_j= 2p\left(p+1\right)}$ data symbols employing ACS and $T=2p+1$ symbol extensions in time.

% \newpage

%%%%%%%%%%%%%%%%%%%%%%%%%%%%%%%%%%%%%%%%%%%%%%%%%%%%%%%%%
%%%%%%%%%%%%%%%%%%%%%%%%%%%%%%%%%%%%%%%%%%%%%%%%%%%%%%%%%
%%%%%%%%%%%%%%%%%%%%%%%%%%%%%%%%%%%%%%%%%%%%%%%%%%%%%%%%%

\section{Precoding matrix structure}
\label{sec:PrecodingStructure} 
The \textit{subspace alignment chains} concept \cite{AlignmentChainsTrans} describes a linear precoding strategy whereby the transmit precoders of the different users are connected for getting the alignment of interfering signals at each receiver. For the proper alignment of interfering signals at the receivers, the precoding matrix of each user is divided in $p$ sub-block matrices, grouped in three main matrix blocks,
\begin{IEEEeqnarray}{l}
\Vn_i \!  = \! \! 
\begin{pmat}[{..|.|..}]
	 \Vn_{i,\left(1\right)}^1 
	& \ldots &
	 \Vn_{i,\left(S_i^1\right)}^1 
	& 
	 \Vn_{i,\left(1\right)}^2 
	&  \ldots 
	& \ldots &
	 \Vn_{i,\left(S_i^3 \right)}^3 
	\cr
\end{pmat}
\mathbf{P}_i,
\label{eq:precMatrix}
\end{IEEEeqnarray}
where $\mathbf{P}_i \Rmat{\hat{d}_j}{\hat{d}_j}$ is an arbitrary unitary permutation matrix used to obtain the same structure for all users and $\Vn_{i,\left(s\right)}^{k} \Rmat{2Tp}{2(p+1)}$ denotes the $s$th sub-block of the $i$th user designed according to the $k$th alignment chain condition. Three alignment chains are built, describing the constraints to be satisfied by each sub-block, see (\ref{eq:AlternateAlignChaina})-(\ref{eq:AlternateAlignChaind}), where $k=1,2,3$ identifies each alignment chain, $\eta_k = k - p$ is the last receiver of the $k$th chain, and the value $S_i^k$ denotes the number of sub-blocks corresponding to the $i$th user designed according to the $k$th alignment chain. Since there are 3 users and each precoding matrix has $p$ sub-blocks, $S_i^k$ may be expressed in closed form as
\setcounter{equation}{19}
\begin{IEEEeqnarray}{c}
S_i^k = 
\left\lceil 
	\frac {p - \left\langle {k - i} \right\rangle} {3} 
\right\rceil .
\end{IEEEeqnarray}

\begin{figure*}
\setcounter{equation}{14}
\begin{IEEEeqnarray}{c}
\Span{ \Hn_{k+1,k}{\Vn_{k,(1)}^k }} = 
\Span{ \Hn_{k+1,k-1}\mathbf{V}_{k-1,(1)}^k}   
\label{eq:AlternateAlignChaina} \\
\Span{ \Hn_{k,k-1}{\Vn_{k-1,(1)}^k }} = 
\Span{ \Hn_{k,k+1}\mathbf{V}_{k+1,(1)}^k}   
\label{eq:AlternateAlignChainb} \\
\Span{ \Hn_{k-1,k+1}{\Vn_{k+1,(1)}^k }} = 
\Span{ \Hn_{k-1,k}\mathbf{V}_{k,(2)}^k}   
\label{eq:AlternateAlignChainc} \\
\nonumber \vdots \\
\Span{ \Hn_{\eta_k,\eta_k-1}
\Vn_{\eta_k-1,\big(S_{\eta_k-1}^k\big)}^k } = 
\Span{ \Hn_{\eta_k,\eta_k+1} \Vn_{\eta_k+1,\big(S_{\eta_k+1}^k\big)}^k }
\label{eq:AlternateAlignChaind}  
\end{IEEEeqnarray} 
% \vspace{-2mm}
\begin{IEEEeqnarray}{c}
\begin{bmatrix}
	\Hn_{k+1,k} & -\Hn_{k+1,k-1} & \0 & \ldots & \0 \\
	\0          &\Hn_{k,k-1}    & -\Hn_{k,k-2} & &\vdots \\
	\vdots & \ddots & \ddots & \ddots & \0 \\
 	\0 & \ldots & \0 & \Hn_{\eta_k,\eta_k-1}
	&-\Hn_{\eta_k,\eta_k+1}
\end{bmatrix}
\begin{bmatrix}
	\Vn_{k,\left(1\right)}^k \\
	\Vn_{k-1,\left(1\right)}^k \\
	\Vn_{k-2,\left(1\right)}^k \\
	\Vn_{k,\left(2\right)}^k \\
	\vdots \\
	\Vn_{\eta_k+1,\left(S_{\eta_k+1}^k\right)}^k
\end{bmatrix} = \0
\label{eq:GeneralAlignChain}
\end{IEEEeqnarray}

\hrule
\vspace*{-8mm}
\end{figure*}

\begin{figure}[]
% defining custom colors
\begin{center} 
  \centerline{\includegraphics[width=0.45\linewidth]{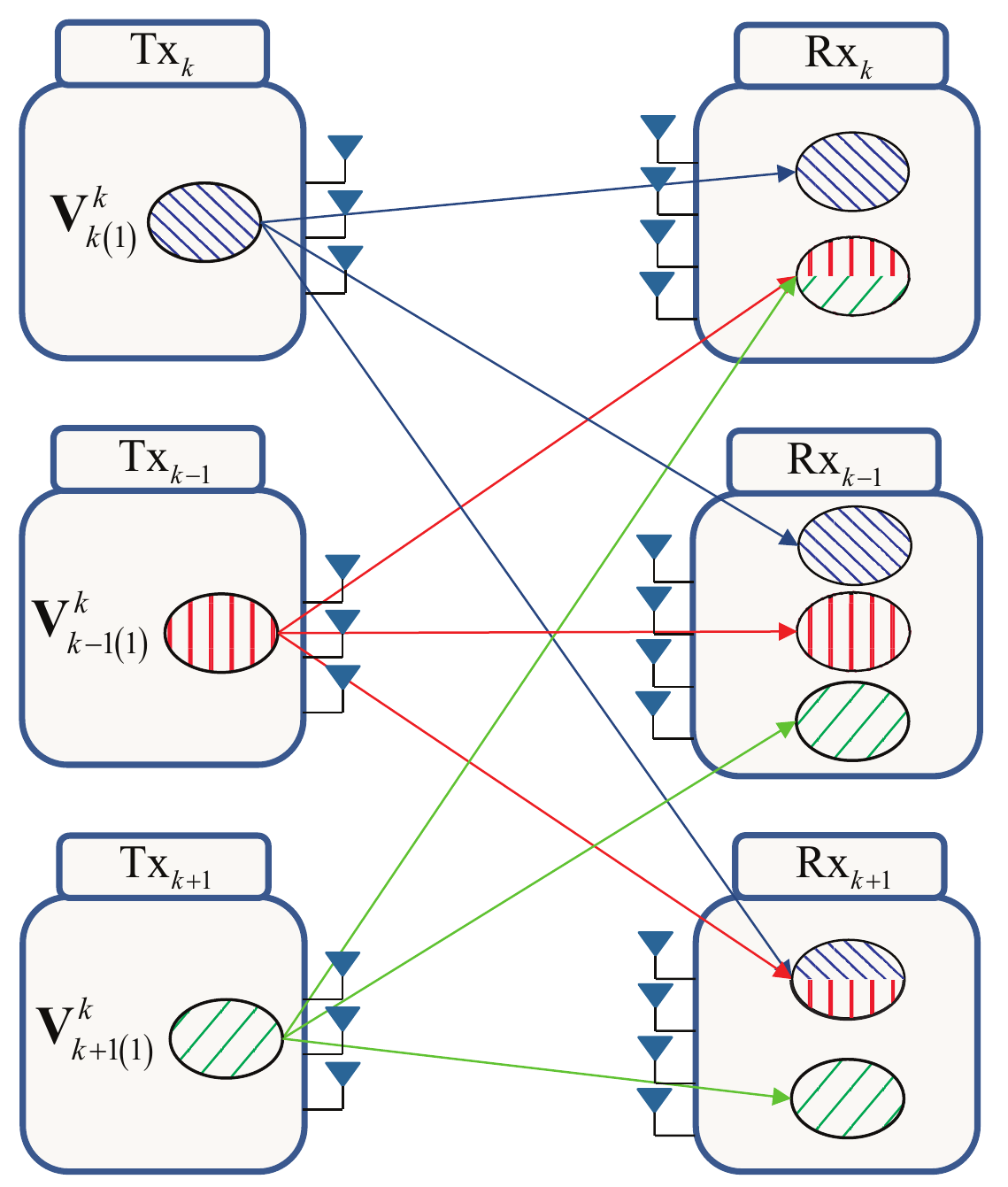}}   
\caption[General Alignment chain]
{Occupation of receivers for the signals designed using alignment chain $k$ ($p=3$ case). Ovals represent different subspaces at transmitters and receivers. Colors/Line patterns identify users.}
\label{fig:alignChainNoSquare}
\end{center} 
\vspace{-6mm}
\end{figure}

The meaning of the alignment chain conditions in (\ref{eq:AlternateAlignChaina})-(\ref{eq:AlternateAlignChaind}) is reviewed in the sequel, and depicted for the $p=3$ case in Fig. \ref{fig:alignChainNoSquare}, where ovals represent the subspaces for the $k$th alignment chain at each transmitter/receiver, and each color/line pattern identifies each user's signals. First, (\ref{eq:AlternateAlignChaina}) states that the subspace occupied by the sub-block $\mathbf{V}_{k,(1)}^k $ should be same as that for the sub-block $\mathbf{V}_{k-1,(1)}^k$ at the $(k+1)$th receiver, see Fig. \ref{fig:alignChainNoSquare}. In the literature, this is usually expressed as the alignment among sub-block $\mathbf{V}_{k,(1)}^k $ and sub-block $\mathbf{V}_{k-1,(1)}^k$ at receiver $(k+1)$. Next, (\ref{eq:AlternateAlignChainb}) ensures that this latter sub-block is, $\textit{at the same time}$, aligned with $\mathbf{V}_{k+1,(1)}^k$ at the $k$th receiver. This process continues as long as there exists a subspace at each receiver where signals can be aligned. The existence of such subspace can be guaranteed by means of basic linear algebra properties (see \cite{AlignmentChainsTrans} for details), and defines the length of the alignment chain, corresponding to the number of sub-blocks designed according to such chain. Notice that the first and last sub-blocks in each alignment chain participate only in the first and the last conditions, respectively. Consequently, they are only aligned with other undesired signals at one of the non-intended receivers. This can be observed in Fig. \ref{fig:alignChainNoSquare} at receiver $k-1$.

The resolution of (\ref{eq:AlternateAlignChaina})-(\ref{eq:AlternateAlignChaind}) is typically tackled by dropping the $\Span{\cdot}$ operators and simply equating the matrices at both sides, as in (\ref{eq:GeneralAlignChain}). Hence, the precoding matrices are obtained as the right null space of some matrix\footnote{For $p=2$ the notation has to be minorly changed. This case will be addressed in Section \ref{sec:5case23}.}. Notice that (\ref{eq:GeneralAlignChain}) represents a sufficient but not necessary condition for (\ref{eq:AlternateAlignChaina})-(\ref{eq:AlternateAlignChaind}). In other words, (\ref{eq:GeneralAlignChain}) is more restrictive than (\ref{eq:AlternateAlignChaina})-(\ref{eq:AlternateAlignChaind}), but it is sufficient for our purpose.

Finally, for convenience in the analysis each sub-block $\Vn_{i,\left(s\right)}^{k} $ is subsequently divided in $p$ blocks by rows, as follows:
\setcounter{equation}{20}
\begin{IEEEeqnarray}{c}
\label{eq:subprecodingMatrix}
{\mathbf{V}}_{i,\left( s \right)}^k = \!
\Stack {
 \Vn_{i,\left(s\right)}^{k,1} ,
\Vn_{i,\left(s\right))}^{k,2} ,
\!  \dots,  \,\,
 \Vn_{i,\left(s\right))}^{k,p} },
\end{IEEEeqnarray}
where each $\Vn_{i,\left(s\right)}^{k,r} \Rmat{2T}{2(p+1)}$ corresponds to one of the $r=1\dots p$ transmit antennas.
%%%%%%%%%%%%%%%%%%%%%%%%%%%%%%%%%%%%%%%%%%%%%%%%%%%%%%%%%
%%%%%%%%%%%%%%%%%%%%%%%%%%%%%%%%%%%%%%%%%%%%%%%%%%%%%%%%%
\section{The $\left(2,3\right)$ Case}
\label{sec:5case23}
This section characterizes the DoF of the (2,3) constant MIMO IC. A precoding scheme is presented where each transmitter delivers $\hat d_j = 12$ real-valued symbols to its intended receiver over $2T=10$ channel extensions, thus attaining the DoF outer bound of $\nicefrac{6}{5}$ according to (\ref{eq:DoFouterbound}). First, the precoding matrices are obtained for this antenna deployment in Section \ref{sec:5A}, designed according to minorly modified conditions from the ones shown in Section \ref{sec:PrecodingStructure}. Next, Section \ref{sec:5B} derives the SSM $\mathbf{G}_j$ introduced in (\ref{eq:Gdefinition}) and provides the achievability proof for the proposed precoding scheme. 

%%%%%%%%%%%%%%%%%%%%%%%%%%%%%%%%%%%%%%%%%%%%%%%%%%%%%%%%%
\def\arraystretch{1}

\subsection{Precoding matrix design}
\label{sec:5A}
According to definitions (\ref{eq:precMatrix}) and (\ref{eq:subprecodingMatrix}), each precoding matrix can be written as
\begin{IEEEeqnarray}{c}
\label{eq:precod_p2}
\Vn_i = {
\begin{bmatrix}
	\Vn_i^{1} & \Vn_i^{2}\\ 
\end{bmatrix}
},
\quad
\Vn_i^k = {
\begin{bmatrix}
	\Vn_i^{k,1} \\ 
	\Vn_i^{k,2} 
\end{bmatrix}
},
\end{IEEEeqnarray}
with ${\mathbf{V}_i\Rmat{20}{12}}$, ${\mathbf{V}_i^k\Rmat{20}{6}}$ and ${\mathbf{V}_i^{k,q}\Rmat{10}{6}}$. Notice that for ease of notation the subindex $s$ appearing in (\ref{eq:precMatrix}) has been dropped. 

The three alignment chains for this case are
\begin{IEEEeqnarray*}{c}
\label{eq:alChains23}
\begin{matrix}
\big[
	\Hn_{2,1},-\Hn_{2,3}
\big] \!\!
\begin{bmatrix}
	\Vn_1^1 \\ \Vn_3^1
\end{bmatrix}
= \0 , \quad
\big[
	\Hn_{3,2} , -\Hn_{3,1}
\big] \!\!
\begin{bmatrix}
	\Vn_2^1 \\ \Vn_1^2
\end{bmatrix}
= \0  , \,
\quad
\big[
	\Hn_{1,3} , -\Hn_{12}
\big] \!\!
\begin{bmatrix}
	\Vn_3^2 \\ \Vn_2^2
\end{bmatrix}
= \0 .
\end{matrix}
\end{IEEEeqnarray*}
Next, we focus without loss of generalization on the first alignment chain. By plugging the particular structure of equivalent channels (see Appendix \ref{sec:appendixA}), it reduces to
\begin{IEEEeqnarray*}{c}
\setlength{\arraycolsep}{2pt}
\begin{bmatrix}
\mathbf{C}\left( {h}_{2,1}^{1,1} \right) & \0 & \0 & \0 \\
\0 & \mathbf{C}\left( {h}_{2,1}^{2,2} \right) 
& \mathbf{C}\left( {h}_{2,3}^{2,1} \right) & \0 \\
\0 & \0 & \0 & \mathbf{C}\left( {h}_{2,3}^{3,2} \right)\\
\end{bmatrix} \!\!
\begin{bmatrix}
{{\mathbf{V}}_1^{1,1}}\\
{{\mathbf{V}}_1^{1,2}}\\
{{\mathbf{V}}_3^{1,1}}\\
{{\mathbf{V}}_3^{1,2}}
\end{bmatrix} = \0 . \nonumber
\label{eq:22_alChains23}
\end{IEEEeqnarray*}
\setlength{\arraycolsep}{5pt} 
\def\arraystretch{1}
This is easily solved by using properties in (\ref{eq:propertiesU}) and taking into account that non-zero blocks are full-rank with high probability, obtaining
\begin{IEEEeqnarray*}{c}
{{\mathbf{V}}_1^{1,1} = {\0}
},
\quad
{{\mathbf{V}}_3^{1,1} = \Cn\left( {\frac{{{h}_{2,1}^{2,2}}}{{{h}_{2,3}^{2,1}}}} \right){\mathbf{V}}_1^{1,2}
},
\quad
{{\mathbf{V}}_3^{1,2} = {\0}}. \nonumber   \label{eq:23solutionAlChains23}
\end{IEEEeqnarray*}
Similarly, one may solve the rest of alignment chains, finally obtaining
\def\arraystretch{1}
\begin{IEEEeqnarray*}{c}
{\mathbf{V}_1} = 
\begin{bmatrix}
\0& \mathbf{C} \left( \frac{{h}_{3,2}^{2,2}}{{h}_{3,1}^{2,1}} \right) \Vn_2^{1,2}\\
\Vn_1^{1,2}&\0
\end{bmatrix} 
\mathbf{P}_1, \qquad
{\mathbf{V}_2} = 
\begin{bmatrix}
\0& \mathbf{C} \left( \frac{{h}_{1,3}^{2,2}}{{h}_{12}^{2,1}} \right) \Vn_3^{2,2}\\
\Vn_2^{1,2}&\0
\end{bmatrix} 
\mathbf{P}_2,
\\
{\mathbf{V}_3} = 
\begin{bmatrix}
\mathbf{C} \left( \frac{{h}_{2,1}^{2,2}}{{h}_{2,3}^{2,1}} \right) \Vn_1^{1,2} & \0\\
\0 & \Vn_3^{2,2}
\end{bmatrix} 
\mathbf{P}_3.
\end{IEEEeqnarray*}

Now we will make use of the permutation matrices $\mathbf{P}_i$ in order to obtain the same structure for all precoding matrices. Notice that reordering the columns of the precoders does not affect to the interference alignment. Furthermore, notice that there are only three precoding sub-blocks different from zero. Hereafter, they will be referred to as the support precoding blocks (SPBs) and denoted as $\An_1,\An_2$ and $\An_3$. Therefore, the $j$th precoding matrix for $j=1,2,3$ is generally written as follows:
\setcounter{equation}{23}
\begin{IEEEeqnarray}{c}
\Vn_j = 
\begin{bmatrix}
\mathbf{C} \left( \frac{{h}_{j - 1,j + 1}^{2,2}}{{h}_{j - 1,j}^{2,1}} \right) \An_{j + 1} & \0 \\
\0 & \An_j
\end{bmatrix}.
\label{eq:25}
\end{IEEEeqnarray}

%%%%%%%%%%%%%%%%%%%%%%%%%%%%%%%%%%%%%%%%%%%%%%%%%%%%%%%%%
\subsection{Achievability proof}
\label{sec:5B}
This section derives the SSM $\mathbf{G}_j$ as a function of the SPBs. Then, a design for those matrices is proposed easing the achievability proof, formalized in \mbox{Lemma \ref{lem:p=2}}.

For the proper computation of the SSM, let write 
\setlength{\arraycolsep}{2pt}
\begin{IEEEeqnarray}{c}
\small
\big[ \Hn_{j,j + 1}\!\Vn_{j + 1},\Hn_{j,j - 1} \Vn_{j - 1} \big] = \!\!
\begin{bmatrix}
\Cn \! \left( \frac{{h}_{j,j - 1}^{1,1} {h}_{j + 1,j}^{2,2}} 
{{h}_{j+1,j-1}^{2,1}} \right) \! \An_j \!\!
& \0 & \0 & \0\\
\0 & \Cn \! \left( {{h}_{j,j - 1}^{2,2}} \right) \! \An_{j - 1}
& \Cn \! \left( {h}_{j,j - 1}^{2,2} \right) \!
\An_{j - 1}\! & \0\\
\0 & \0 & \0 & \Cn \! \left( {{h}_{j,j + 1}^{3,2}} \right) \!
\An_{j + 1}
\end{bmatrix} \nonumber \\
\label{eq:crossProducts}
\end{IEEEeqnarray}
defining the subspaces of received interference at the $j$th receiver, see (\ref{eq:Gdefinition}).
Notice that the third block column of (\ref{eq:crossProducts}) is aligned with the second block column of (\ref{eq:crossProducts}), which is actually forced by the alignment chain $j+1$. As a result, the basis for the interfering space $\Gint$ is defined by the three linearly independent block columns of (\ref{eq:crossProducts}), and the SSM $\mathbf{G}_j$ is given by (\ref{eq:28matrixG_23}).

\def\arraystretch{1}
\setlength{\arraycolsep}{5pt}
\begin{figure*}
\setcounter{equation}{22}
\def\arraystretch{1.3}
\setlength{\arraycolsep}{5pt}
 \begin{IEEEeqnarray}{c}
\Gn_j=
\begin{bmatrix}
\Cn\left( {h}_{j,j}^{1,1} \right) \An_{j + 1} &
\Cn\left( {h}_{j,j}^{1,2} \right) \An_j  &
\Cn\left( \frac{{h}_{j,j - 1}^{1,1} {h}_{j + 1,j}^{2,2}}
{{h}_{j+1,j-1}^{2,1}} \right) \An_j &
\0 & \0
\\
\Cn\left( {h}_{j,j}^{2,1} \right) \An_{j + 1} &
\Cn\left( {h}_{j,j}^{2,2} \right) \An_j &
\0 &
\Cn \left( {h}_{j,j - 1}^{2,2} \right) \An_{j-1} &
\0
\\[2mm]
\Cn \left( {h}_{j,j}^{3,1} \right) \An_{j+1} &
\Cn \left( {h}_{j,j}^{3,2} \right) \An_j &
\0 &
\0 &
\Cn \left( {h}_{j,j + 1}^{3,2} \right) \An_{j + 1}
\end{bmatrix}
\def\arraystretch{1} \nonumber \\
 \label{eq:28matrixG_23}
\end{IEEEeqnarray}
\hrule
\vspace{-10mm}
\end{figure*}

The SSM obtained in (\ref{eq:28matrixG_23}) is similar to the equivalent magnitude obtained in equation (16) of \cite{ACSnoPublicat}. Even though in this case the full-rank condition for the SSM can be ensured by picking entries of the SPBs randomly (as pointed out by \cite{ACSnoPublicat}), we present a formal proof that is also useful for the non-straightforward $p>2$ case.
 
Let define $\boldsymbol\lambda_i^j \Rmat{6}{1}$, $i=1\ldots5$, $j=1,2,3$ as the \textit{rank multipliers}. Then, one may ensure that the SSM is full-rank iff the only solution for
\setcounter{equation}{25}
\begin{IEEEeqnarray}{c}
\label{eq:29Glanda}
\Gn_j 
\Big[
\, \left( \boldsymbol{\lambda }_1^j \right)^T \,\ldots \, \left(\boldsymbol{\lambda }_5^j\right)^T \,\,
\Big]^T
= \0
\end{IEEEeqnarray}
is to set all rank multipliers to zero. To this end, let also define an arbitrary orthonormal basis ${\mathbf{B}= \Big[ 
\mathbf{b}_{1} \, \mathbf{b}_{2} \, \ldots \, \mathbf{b}_{10} \Big] \Rmat{10}{10}}$. We propose the following design: 
\begin{IEEEeqnarray}{c}
\setlength{\arraycolsep}{6pt}
\begin{matrix}
\An_1 = 
\begin{bmatrix}
	\mathbf{B}_{1:2} & \mathbf{B}_{3:5} & \mathbf{B}_6
\end{bmatrix},
\quad
\An_2 = 
\begin{bmatrix}
	\mathbf{B}_{1:2} & \mathbf{B}_{7:9} & \mathbf{B}_{10}
\end{bmatrix},
\quad
\An_3 = 
\begin{bmatrix}
	\mathbf{B}_{3:5} & \mathbf{B}_{7:9}
\end{bmatrix}.
\end{matrix}
\label{eq:30Adesign}
\end{IEEEeqnarray}
The following lemma states the DoF achievability:
\begin{lemma}[$\mathbf{G}_j$ full-rank for $p=2$]
\label{lem:p=2}
Considering (\ref{eq:28matrixG_23}) and the SPBs chosen as in (\ref{eq:30Adesign}), then the only possible solution for (\ref{eq:29Glanda}) is $\boldsymbol{\lambda}_i^j=\0, \forall i,j$.
\end{lemma}
\begin{IEEEproof}
See Appendix \ref{sec:appendixB}.
\end{IEEEproof}

Finally, the optimal DoF are settled by means of the following theorem:
\begin{theorem}[DoF for the (2,3) case]
\label{theorem23}
The 3-user $\left(2,3 \right)$ MIMO IC with constant channel coefficients has exactly 6/5 DoF per user, and they can be achieved by means of linear precoding at the transmitters and linear filtering at the receivers.
\end{theorem}
\begin{IEEEproof}
Each user transmits $ \hat{d}=12$ real-valued symbol streams along $T=5$ symbol extensions in time, considering ACS, and the precoding scheme described in Section \ref{sec:5A}. Therefore, according to Lemma \ref{lem:p=2}, the SSM $\mathbf{G}_j$ becomes full rank, thus interference and desired signals become linearly independent, and the desired symbols can be decoded. Since the DoF outer bound (\ref{eq:DoFouterbound}) and the achievable DoF attained by the proposed scheme match, this value corresponds to the optimal DoF.
\end{IEEEproof}

%%%%%%%%%%%%%%%%%%%%%%%%%%%%%%%%%%%%%%%%%%%%%%%%%%%%%%%%%
%%%%%%%%%%%%%%%%%%%%%%%%%%%%%%%%%%%%%%%%%%%%%%%%%%%%%%%%%
\section{The $\left(p,p+1\right)$ case with $p>2$}
\label{sec:6generalCase}
This section defines the optimal DoF for the $p\geq 3$ case. A precoding scheme is presented where each user obtains $\hat d_j = 2p\left(p+1\right)$ real-valued data symbols over $2T=2\left(2p+1\right)$ channel extensions, thus attaining the DoF outer bound $\frac{p\left(p+1\right)}{2p+1}$ in (\ref{eq:DoFouterbound}).

Unfortunately, the number of conditions used for the precoder design, see (\ref{eq:GeneralAlignChain}), increases with $p^2$. Therefore, the complexity of the analysis using the approach for the $p=2$ case becomes cumbersome as $p$ grows. This section presents a methodology to simplify the resolution of such matrix equation system, which will be illustrated for the $p=3$ case. The core of this methodology is the \ac{ZP} algorithm, which allows to obtain the structure of the transmit and receive filters for any value of $p$.

%%%%%%%%%%%%%%%%%%%%%%%%%%%%%%%%%%%%%%%%%%%%%%%%%%%%%%%%%
\subsection{Precoding matrix design}
\label{sec:6A}
Consider the first alignment chain ($k=1$) given by (\ref{eq:31}), shown at the top of the next page. 
\begin{figure*}[t]
\setcounter{equation}{29}
\begin{IEEEeqnarray*}{c}
\def\arraystretch{1.5}
\setlength{\arraycolsep}{2pt}
\begin{bmatrix}
	\Cn \left( {h}_{2,1}^{1,1} \right) 
	\!&\!\0&\0&\0&\0&\0&\0&\0&\0
\\
	\0&\Cn \left( {h}_{2,1}^{2,2} \right)&
	\Cn \left( {h}_{2,1}^{2,3} \right)&
	\Cn \left( {h}_{2,3}^{2,1} \right)&\0&\0&\0&\0&\0 
\\ \0 & \0 & \Cn \left( {h}_{2,1}^{3,3} \right)&
	\Cn \left( {h}_{2,3}^{3,1} \right)&
	\Cn \left( {h}_{2,3}^{3,2} \right) & \0&\0&\0&\0
\\
	\0&\0&\0&\0&\0&\Cn \left( {h}_{2,3}^{4,3} \right)&\0&\0&\0
\\
	\0&\0&\0&\Cn \left( {h}_{1,3}^{1,1} \right) &\0&\0&\0&\0&\0
\\
	\0&\0&\0&\0&\Cn \left( {h}_{1,3}^{2,2} \right)&
	\Cn \left( {h}_{1,3}^{2,3} \right)&
	\Cn \left( {h}_{1,2}^{2,1} \right)&\0&\0 
\\ \0&\0&\0 & \0 & \0 & \Cn \left( {h}_{1,3}^{3,3} \right)&
	\Cn \left( {h}_{1,2}^{3,1} \right)&
	\Cn \left( {h}_{1,2}^{3,2} \right) & \0
\\
	\0&\0&\0&\0&\0&\0&\0&\0&\Cn \left( {h}_{1,2}^{4,3} \right)
\end{bmatrix}
\begin{bmatrix}
\Vn_{1,\left(1\right)}^{1,1} \\
\Vn_{1,\left(1\right)}^{1,2} \\
\Vn_{1,\left(1\right)}^{1,3} \\
\Vn_{3,\left(1\right)}^{1,1} \\
\Vn_{3,\left(1\right)}^{1,2} \\
\Vn_{3,\left(1\right)}^{1,3} \\
\Vn_{2,\left(1\right)}^{1,1} \\
\Vn_{2,\left(1\right)}^{1,2} \\
\Vn_{2,\left(1\right)}^{1,3} \\
\end{bmatrix} 
= \0 
\end{IEEEeqnarray*}
\begin{align}
\label{eq:31}
\mathbf{E} \cdot \mathbf{F} = \0
\end{align}
\vspace{-8mm}
\centering
\def\arraystretch{1.2}
\setlength{\arraycolsep}{6pt}
 \begin{IEEEeqnarray}{c}
\begin{bmatrix}
	\0 
	&\0&\0&\0&\0&\0&\0&\0&\0
\\
	\0&\Cn \left( {h}_{2,1}^{2,2} \right)&
	\Cn \left( {h}_{2,1}^{2,3} \right)&
	\0&\0&\0&\0&\0&\0 
\\ \0 & \0 & \Cn \left( {h}_{2,1}^{3,3} \right)&
	\0&
	\Cn \left( {h}_{2,3}^{3,2} \right) & \0&\0&\0&\0
\\
	\0&\0&\0&\0&\0&\0&\0&\0&\0
\\
	\0&\0&\0&\0 &\0&\0&\0&\0&\0
\\
	\0&\0&\0&\0&\Cn \left( {h}_{1,3}^{2,2} \right)&
	\0&
	\Cn \left( {h}_{1,2}^{2,1} \right)&\0&\0 
\\ \0&\0&\0 & \0 & \0 & \0&
	\Cn \left( {h}_{1,2}^{3,1} \right)&
	\Cn \left( {h}_{1,2}^{3,2} \right) & \0
\\
	\0&\0&\0&\0&\0&\0&\0&\0&\0
\end{bmatrix}
\begin{bmatrix}
\0 \\
\Vn_{1,\left(1\right)}^{1,2} \\
\Vn_{1,\left(1\right)}^{1,3} \\
\0 \\
\Vn_{3,\left(1\right)}^{1,2} \\
\0 \\
\Vn_{2,\left(1\right)}^{1,1} \\
\Vn_{2,\left(1\right)}^{1,2} \\
\0 \\
\end{bmatrix} 
= \0 
 \label{eq:outputZP}
 \setlength{\arraycolsep}{5pt}
\end{IEEEeqnarray}
\\
\hrule

\vspace{-12mm}
\end{figure*}

It can be observed that thanks to the obtained structure of matrix $\mathbf{E}$, some sub-blocks of $\mathbf{F}$ are zero. For example, consider the fifth block row element
\setcounter{equation}{27}
\begin{IEEEeqnarray}{c}
\Cn\left( {{h}_{1,3}^{1,1}} \right){\mathbf{V}}_{3,\left( 1 \right)}^{1,1}  = \0.
\label{eq:3,2a}
\end{IEEEeqnarray}
Clearly, the only solution for (\ref{eq:3,2a}) is $\mathbf{V}_{3,(1)}^{1,1}=\0$. Hence, other equations where this variable participates are simplified. These events are denoted as {\it zero propagations} (ZP) and give the possibility of finding which blocks are zero for $\mathbf{F}$ in (\ref{eq:31}). Inspired by this idea, we present the ZP algorithm, see \mbox{Table 1}.

\def\arraystretch{1.7}
\begin{table*}
\begin{center}
 \begin{minipage}{0.9\linewidth}
\normalsize
\caption{\small Zero Propagation Algorithm}\label{ZPalg}
\vspace{-6mm}
 \begin{tabular}{l p{0.8\linewidth}}
 \hhline{==} 
 \multicolumn{2}{p{0.94\linewidth}}{Consider the matrix equation system given by ${\mathbf{E} \cdot \mathbf{F} = \0}$,
with ${\mathbf{F} \Rmat{F_\text{BR} \cdot r_\text{F}}{F_\text{BC}}}$ and ${\mathbf{E} \Rmat{E_\text{BR} \cdot r_\text{E}}{F_\text{BR} \cdot r_\text{F}}}$, where $r_\text{F} (r_\text{E})$ defines the number of block rows of $\mathbf{E}$ ($\mathbf{F}$). Moreover, $F_\text{BC}$ ($F_\text{BR}$) defines the number of columns (rows) of each block element of $\mathbf{F}$, and $E_\text{BR}$ defines the number of rows of each block column of $\mathbf{E}$. The blocks of $\mathbf{F}$ that can be set to zero may be obtained by computing the following steps:} \\[-1mm]
1. & Find one block row in $\mathbf{E}$ containing only one non-zero element, located at the $[r^*,c^*]$th block position. \\
2. & Set  $\begin{cases}\mathbf{E}(r^*,:)=\text{zeros}(E_\text{BR}, F_\text{BR} \cdot r_\text{F}) \\   {\mathbf{E}(:,c^*)=\text{zeros}(E_\text{BR} \cdot r_\text{E},F_\text{BR})}\end{cases}$  \\
3. & Set $\mathbf{F}(c^*,:)=\text{zeros}(F_\text{BR},F_\text{BC})$. \\
4. &  Repeat (1)-(3) until (1) provides no more block rows. \\
\hhline{==}
\end{tabular}
\end{minipage}
\end{center}
\vspace{-10mm}
\end{table*} 
\def\arraystretch{1}

This algorithm allows to simplify the conditions initially presented in (\ref{eq:31}) to obtain (\ref{eq:outputZP}). Note that the 1st, 4th, 6th and 9th block elements of $\mathbf{F}$ in (\ref{eq:31}) are forced to be zero. Moreover, by writing the remaining equations it turns out that each precoding matrix can be written as a function of three SPBs, as follows:
\setlength{\arraycolsep}{5pt}
\begin{IEEEeqnarray}{c}
{{\mathbf{V}}_i} = \!
\begin{bmatrix}
 \Cn\left(\theta_{i,(1)}^{i-1,1}\right)\An_{i-1} 
& \0 & \0 
\\
 \Cn\left(\theta_{i,(1)}^{i-1,2}\right)\An_{i-1} 
& \Cn\left(\theta_{i,(1)}^{i+1,2}\right)\An_{i+1} 
& \Cn\left(\theta_{i,(1)}^{i,2}\right)\An_{i} 
\\
\0 & \0 & \An_i
\end{bmatrix},
\label{eq:34} 
\end{IEEEeqnarray}
where $\theta _{i,\left( 1 \right)}^{q,r}$ stands for the complex value obtained from the $q$th alignment chain and located at the $r$th block row of ${\mathbf{V}}_i$. 
Those complex numbers can be obtained by removing the rows and columns with zeros from (\ref{eq:outputZP}) and computing a null space basis.
% Their values can be obtained by removing the empty columns and rows from (\ref{eq:outputZP}) and obtaining a basis for the null space. For our purposes, it is not relevant the analytical expression of $\theta _{i,\left( 1 \right)}^{q,r}$, and we only assume that they maintain the same randomness properties as the original channel coefficients.
Note that the number of unknown sub-block matrices is reduced from 27 in (\ref{eq:31}) to 3 in (\ref{eq:34}). In general, the $3p^2$ variables (block matrices) involved in all alignment chains can be written as a function of the three SPBs of dimension $2(2p+1)\times 2(p+1)$.

%%%%%%%%%%%%%%%%%%%%%%%%%%%%%%%%%%%%%%%%%%%%%%%%%%%%%%%%%
\subsection{Achievability proof}
\label{sec:6B}
This section derives the SSM for the $p=3$ case, and gives some intuitions about the general case. First, a design for the three SPBs in (\ref{eq:34}) is proposed, generalizing (\ref{eq:30Adesign}) for any value of $p$. Second, the SSM is shown to be full rank, hence the optimal DoF are stated in \mbox{Theorem \ref{theorem34}}.

In order to build $\mathbf{G}_j$, it is necessary to compute a basis for the sum space defined by the received interference and desired signals. Regarding the desired signals, it can be easily seen that $\Gn_j^{\text{des}}={{\mathbf{H}}_{j,j}}{{\mathbf{V}}_{j}}$. 
On the other hand, since some of the interference is aligned it is necessary to first calculate the products ${{\mathbf{H}}_{j,j-1}}{{\mathbf{V}}_{j-1}}$ and ${{\mathbf{H}}_{j,j+1}}{{\mathbf{V}}_{j+1}}$. Next, we will see that this task can be highly alleviated. Recall on the fact that the ZP algorithm output in (\ref{eq:outputZP}) not only states which sub-blocks of each $\Vn_i$ are actually zero, but also which conditions should satisfy the remaining sub-blocks. For example, from (\ref{eq:outputZP}) it can be observed that
\vspace{-1mm}
\setcounter{equation}{31}
\begin{IEEEeqnarray}{c}
\Cn\left( {{h}_{2,1}^{2,2}} \right){\mathbf{V}}_{1,\left( 1 \right)}^{1,2} + \Cn\left( {{h}_{2,1}^{2,3}} \right){\mathbf{V}}_{1,\left( 1 \right)}^{1,3} = {\0}
\label{eq:exampleSimpG}
\vspace{-1mm}
\end{IEEEeqnarray}
needs to be satisfied. Interestingly, this is indeed one of the elements resulting from the product $\Hn_{2,1}\Vn_1$. Taking into account all other conditions where there are only elements managed by one unique transmitter, the products ${{\mathbf{H}}_{j,j-1}}{{\mathbf{V}}_{j-1}}$ and ${{\mathbf{H}}_{j,j+1}}{{\mathbf{V}}_{j+1}}$ can be further simplified, obtaining (\ref{eq:prodCreuatsa})-(\ref{eq:prodCreuatsb}), where $\bar \theta _{j,i}^{q,r}$ is the corresponding complex number for the ($q$,$r$)th position of ${\mathbf{H}_{j,i} \mathbf{V}_i , i \ne j}$. Note that in this case due to alignment conditions, we will have $\bar \theta _{j,j - 1}^{q,q} = \bar \theta _{j,j + 1}^{q,q - 1}$ with $q=2,3$, i.e. columns 2, 3 of ${\mathbf{H}}_{j,j + 1}{\mathbf{V}}_{j + 1}$ are aligned with columns 1, 2 of ${\mathbf{H}}_{j,j - 1}{\mathbf{V}}_{j - 1}$, respectively. Therefore, in this case the SSM is given by (\ref{eq:Gdes_p3})-(\ref{eq:Gint_p3}), where $\Gdes(q,r)$ and $\hat \theta _j^{q,r}$ are the matrix and the complex number corresponding to the ($q$,$r$)th position of $\Gdes$ and $\Gint$, respectively. For $\Gdes$, we write the blocks $\Gdes(q,r)$ because they are linear combinations of some extended channel elements, e.g. 
\vspace{-1mm}
\begin{IEEEeqnarray*}{c}
{\Gdes(1,2)=\Ch{1,1}{1}}-\C{ \frac{\hel{1,1}{2,2}\hel{3,1}{3,1}}{\hel{3,1}{3,2}} }.
\vspace{-1mm}
\end{IEEEeqnarray*}
Notice that each matrix $\Cn \big(\hat \theta _j^{q,r}\big)$ is a combination of a number of cross-channels coefficients, thus it can be assumed independent of any of the matrices $\Gdes(q,r)$, since they are function of the direct channel coefficients. 

\begin{figure*}[t]
\setcounter{equation}{35}
\def\arraystretch{1.6}
 \begin{IEEEeqnarray}{r l}
\Hn_{j,j-1} \Vn_{j-1} = &
\begin{bmatrix}
  \Cn\left(\bar\theta_{j,j-1}^{1,1}\right)\An_{j+1}
& \0 & \0
\\
  \Cn\left(\bar\theta_{j,j-1}^{2,1}\right)\An_{j+1}
& \Cn\left(\bar\theta_{j,j-1}^{2,2}\right)\An_{j}
& \0
\\
  \0 & \0
& \Cn\left(\bar\theta_{j,j-1}^{3,3}\right)\An_{j-1}
\\
  \0 & \0 & \0
\end{bmatrix} 
\label{eq:prodCreuatsa} \\[2mm]
\Hn_{j,j+1} \Vn_{j+1} = &
\begin{bmatrix}
  \0 & \0 & \0
\\
  \Cn\left(\bar\theta_{j,j+1}^{2,1}\right)\An_{j}
& \0 & \0
\\
  \0 
& \Cn\left(\bar\theta_{j,j+1}^{3,2}\right)\An_{j-1}
& \Cn\left(\bar\theta_{j,j+1}^{3,3}\right)\An_{j+1}
\\
  \0 & \0
& \Cn\left(\bar\theta_{j,j+1}^{4,3}\right)\An_{j+1}
\end{bmatrix}
\label{eq:prodCreuatsb} \\[2mm]
\Gn_j^{\text{des}} =  &
\begin{bmatrix}
  \Geldes{1,1}{j-1}\!\! & \Geldes{1,2}{j+1} & \Geldes{1,3}{j}
\\
  \Geldes{2,1}{j-1}\!\! & \Geldes{2,2}{j+1} & \Geldes{2,3}{j}
\\
  \Geldes{3,1}{j-1}\!\! & \Geldes{3,2}{j+1} & \Geldes{3,3}{j}
\\
  \Geldes{4,1}{j-1}\!\! & \Geldes{4,2}{j+1} & \Geldes{4,3}{j}
\end{bmatrix}
\label{eq:Gdes_p3} \\[2mm]
\Gn_j^{\text{int}} = &
\begin{bmatrix}
  \Gel{1,1}{j+1}\! & \G0 & \G0 & \G0
\\
  \Gel{2,1}{j+1}\! & \Gel{2,2}{j} & \G0 & \G0
\\
  \G0 & \G0\! & \Gel{2,3}{j-1} & \Gel{2,4}{j+1}
\\
  \G0 & \G0\! & \G0 & \Gel{3,4}{j+1}
\end{bmatrix}
\label{eq:Gint_p3}
\end{IEEEeqnarray}
\hrule
\end{figure*}

%The reader can refer to Appendix \ref{sec:appendixC2} to find the key ideas to generalize the structure of matrix $\mathbf{G}_j$ for any value of $p$ using the results for $p=2,3,\ldots,6$. 

In contrast to (\ref{eq:28matrixG_23}), now it is not that clear if the SSM for this case is full-rank by just taking the SPBs randomly. Next, we provide the proof to verify that $\Gn_j$ is full rank. Magnitudes are defined for a general value of $p$, and all possible procedures are generalized. 
As before, the SSM may be shown to be full rank iff all
${\boldsymbol\lambda_i^j \Rmat{2(2p+1)}{1}}$, $i=1\ldots 2p+1$, $j=1,2,3$ constrained by
\setcounter{equation}{32}
\begin{IEEEeqnarray}{c}
\label{eq:29GlandaXX}
\Gn_j 
\Big[
\, \left( \boldsymbol{\lambda }_1^j \right)^T \,\ldots \, \left(\boldsymbol{\lambda }_{2p+1}^j\right)^T \,\,
\Big]^T
= \0
\end{IEEEeqnarray}
are actually equal to zero. 
Define an orthonormal basis $\mathbf{B}=\!\! \left[ \mathbf{b}_1, \mathbf{b}_2\ldots \! \mathbf{b}_{2(2p+1)} \right] \!
\Rmat{2(2p+1)}{2(2p+1)}$ and sets
\begin{IEEEeqnarray}{c}
\begin{matrix}
X_1 = \left\{ 3,4,\ldots,p+3 \right\}, \quad
X_2 = \left\{ p+4,p+5,\ldots,2p+2\right\}, \\
Y_1 = \left\{ 2p+3,\ldots,3p+3\right\}, \quad
Y_2 = \left\{ 3p+4,\ldots,4p+2\right\}, \quad
Z = \left\{1,2\right\}.
\end{matrix}
\label{eq:sets}
\end{IEEEeqnarray}
We will use these sets to arrange columns of $\mathbf{B}$, e.g. $\mathbf{B}_{X_2}=\mathbf{B}_{p+4:2p+2}$. Accordingly, we set:
\begin{IEEEeqnarray}{c}
\begin{matrix}
\An_1 =
\begin{bmatrix}
	\mathbf{B}_Z & \mathbf{B}_{X_1} & \mathbf{B}_{X_2}
\end{bmatrix}, \quad
\An_2 =
\begin{bmatrix}
	\mathbf{B}_Z & \mathbf{B}_{Y_1} & \mathbf{B}_{Y_2}
\end{bmatrix},
\quad
\An_3 =
\begin{bmatrix}
	\mathbf{B}_{X_1} & \mathbf{B}_{Y_1}
\end{bmatrix}.
\end{matrix}
\label{eq:37}
\end{IEEEeqnarray}
Given these definitions, the following lemma states the DoF achievability:
\vskip 2ex 

\begin{lemma}[$\mathbf{G}_j$ full-rank for $p=3\ldots6$]
\label{lem:p=3}
For the $p=3, \ldots, 6$ cases, the SSM defined as in (\ref{eq:Gdefinition}) with SPBs chosen as in (\ref{eq:37}) is full rank with probability one.
\end{lemma}
\begin{IEEEproof}
See Appendix \ref{sec:appendixB1}.
\end{IEEEproof}
\vskip 2ex 

The DoF characterization for $p=3\ldots6$ follows from Lemma 2, and it is next formalized:

\begin{theorem} [DoF of the $\left(p,p+\!1\right)$\! IC,\! $p=3\ldots6$]
\label{theorem34}
The 3-user $\left(p,p+1\right)$, $p=3\ldots6$ MIMO IC with constant channel coefficients has exactly $\frac{p\left(p+1\right)}{2p+1}$ DoF per user, and they can be achieved by means of linear precoding at the transmitters and linear filtering at the receivers.
\end{theorem}

\begin{IEEEproof}
The proof is analogous to the proof of Theorem \ref{theorem23}. In general, the optimal DoF are attained by using the proposed transmitting scheme, delivering $\hat d_j= 2p(p+1)$ symbol streams to each user along $2T=2(2p+1)$ symbol extensions in time, and considering ACS.
\end{IEEEproof}
\vskip 2ex 

We remark that we have only analytically proved the cases $p=2,3,\ldots,6$. Nonetheless, based on the explained methodology and some numerical results (see next section), we conjecture that for any $p>6$ full rank SSMs are obtained, and hence the optimal DoF can be attained:

\begin{conjecture}[DoF for the general $\left(p,p+1\right)$ IC]
\label{conjecture}
The 3-user $\left(p,p+1\right)$ MIMO IC with constant channel coefficients has exactly $\frac{p\left(p+1\right)}{2p+1}$ DoF per user for $p>6$. They can be achieved using linear transmit and receive filters, and by means of applying subspace alignments chains, symbol extensions in time and ACS.
\end{conjecture}

%%%%%%%%%%%%%%%%%%%%%%%%%%%%%%%%%%%%%%%%%%%%%%%%%%%%%%%%%%%%%%%%%%%%%%%%%%%%%%%%%%%%%%%%%%%%%%%%%%%%%%%%%%%%%%%%%%%%%%%%%%%%%%%%%%%%%%%%%%%%%%%%%%%%%%%%%%%%%%%%%%%%%%%%%%%
%%%%%%%%%%%%%%%%%%%%%%%%%%%%%%%%%%%%%%%%%%%%%%%%%%%%%%%%%%%%%%%%%%%%%%%%%%%%%%%%%%%%%%%%%%%%%%%%%%%%%%%%%%%%%%%%%%%%%%%%%%%%%%%%%%%%%%%%%%%%%%%%%%%%%%%%%%%%%%%%%%%%%%%%%%%
%%%%%%%%%%%%%%%%%%%%%%%%%%%%%%%%%%%%%%%%%%%%%%%%%%%%%%%%%%%%%%%%%%%%%%%%%%%%%%%%%%%%%%%%%%%%%%%%%%%%%%%%%%%%%%%%%%%%%%%%%%%%%%%%%%%%%%%%%%%%%%%%%%%%%%%%%%%%%%%%%%%%%%%%%%%

\section{Simulation results}
\label{sec:4.3sim}
\begin{figure}[]
\centering
\footnotesize
\begin{tikzpicture}
\begin{axis}[%
view={0}{90},
width=0.6\linewidth,
xmin=0, xmax=100,
xlabel={SNR(dB)},
grid style={dashed, gray!50},
xmajorgrids,
ymin=0, ymax=400,
ylabel={Sum-rate (bps/Hz)},
ymajorgrids,
ylabel style={at={(0.03,0.5)},rotate=0},
legend style={at={(axis cs: 10,350)},anchor=north west},
cycle multi list={ solid,dashed \nextlist
    {red,mark=square},{blue,mark=diamond},{orange,mark=o},{magenta,mark=triangle},{black,mark=triangle,mark options={rotate=180,solid}},{cyan,mark=pentagon} },    
    every axis plot/.append style={mark size = 2pt,mark repeat={10},mark options={solid}},
    ]
    
    \pgfplotstableread[col sep=tab]{simResults.txt} \rates

    %%%%%%%%%%%%%% ACS %%%%%%%%%%%%%%%%%%%%%%%%%%%%%%%%%%%%%%%%%%%%%%%%
    
    \foreach \L in {2,3,5,6,8,9} {
      \addplot table[x =SNR, y = p\L] from \rates ;
      \addlegendentryexpanded{$p=\L$};
  }

   %%%%%%%%%%%%%% NO ACS %%%%%%%%%%%%%%%%%%%%%%%%%%%%%%%%%%%%%%%%%%%
   
   \foreach \L in {2,3,5,6,8,9} {
      \addplot table[x =SNR, y = np\L] from \rates ;
  }

\end{axis}
\end{tikzpicture}%
\vspace{-2mm}
\caption[]{Comparison of using IA with the proposed channel extension (solid lines) with respect to using the scheme with only symbol extensions in time (dashed lines).}
\label{fig:ACSvsNoACS}
\end{figure}
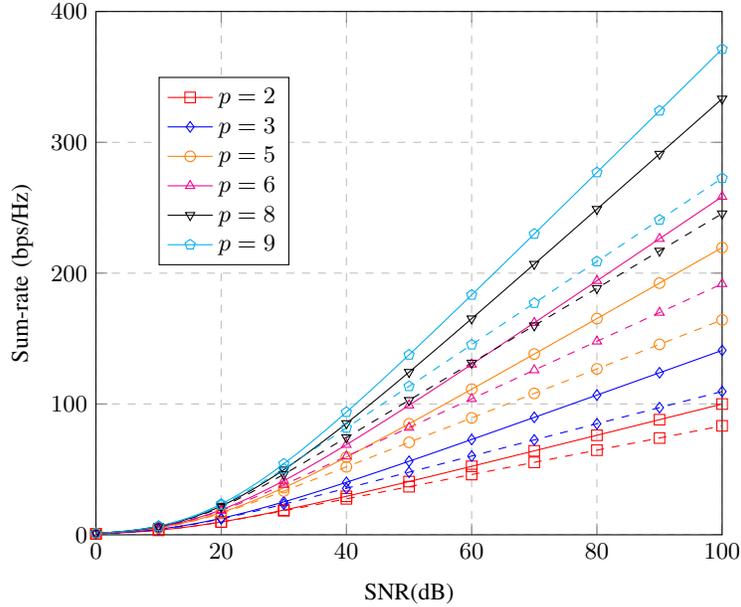

In order to validate the contributions of this work, as well as increase the strength of \mbox{Conjecture \ref{conjecture}}, we simulate the cases $p=2,3,5,6,8,9$ for the 3-user MIMO IC. Two schemes are simulated, the one proposed in this work, and the design in \cite{AlignmentChainsTrans} not considering ACS. In both cases, we apply the CB operation and the additional transformations as explained in Appendix \ref{sec:appendixA} together with the proposed scheme. Results are shown in Fig. \ref{fig:ACSvsNoACS}, where solid/dashed lines denote the two schemes with/without considering ACS. It can be seen that the scheme considering ACS improves the slope achieved at high SNR for each case. Moreover, notice that we simulate two cases $p>6$, whose DoF were conjectured in previous section.

\section{Conclusions} 
\label{sec:7conclusion}
This work has investigated the DoF of the 3-user $(p,p+1)$ MIMO Interference channel with constant channel coefficients and full CSI at both sides. We have obtained that the best known outer bound can be attained for the cases $p=2\ldots6$ by means of a linear precoding scheme. Moreover, a methodology has been presented easing the proof for the general case, where we conjecture that the known DoF outer bound is also tight. This conjecture has been numerically checked for two cases.

The contribution of this work is twofold. On the one hand, we have shown that the use of \ac{ACS} together with the previous state-of-the-art approach in \cite{AlignmentChainsTrans} allows to attain the optimal DoF. Therefore, we have provided a formal proof, and uncoupled the achievability statement from numerical experiments. On the other hand, we have shown that linear precoding schemes attain the same DoF as lattice alignment based schemes (except for the SISO case), being the former more robust for the finite SNR regime. 

Future work may be oriented to complete the characterization of this channel for the SISO case, where DoF inner and outer bounds have not yet been found. Also, it may be interesting to optimize not only the slope of the rate curve at the high SNR regime, but also the SNR offset. Further improvement seems to be possible by optimizing the SPBs in terms of the sum rate subject to some transmit power constraint.

\appendices
%%%%%%%%%%%%%%%%%%%%%%%%%%%%%%%%%%%%%%%%%%%%%%%%%%%%%%%%%
%%%%%%%%%%%%%%%%%%%%%%%%%%%%%%%%%%%%%%%%%%%%%%%%%%%%%%%%%
\section{Additional change of basis at the receiver side}
\label{sec:appendixA}

The CB operation \cite{AlignmentChainsTrans} is a tool that provides a predetermined structure for the cross-channel matrices. In particular, it forces zeros at some specific antenna elements. For example, the equivalent cross-channel matrices $\big\{ \mathbf{\tilde{H}}_{j,j-1},\mathbf{\tilde{H}}_{j,j+1} \big\}$ for $p=3$ after performing the original CB described in \cite{AlignmentChainsTrans} are given by (\ref{eq:estructuraPreFinal_H_p=3}). Here we assume that the CB at the receiver $\mathbf{R}_{j}$ is the product of two matrices: the original CB and an additional combining matrix  $\boldsymbol{\Upsilon}_j \Rmat{2T(p+1)}{2T(p+1)}$ such that 
 (\ref{eq:estructuraFinal_H_p=3}) is satisfied. Then, each block row of $\boldsymbol{\Upsilon}_j=\big[ \boldsymbol{\upsilon}_{j,1}^T , \ldots , \boldsymbol{\upsilon}_{j,4}^T \big]^T$ is derived as follows:
 \setcounter{equation}{41}
\begin{IEEEeqnarray}{c}
\begin{matrix} 
 \boldsymbol{\upsilon}_{j,1} = \big[ \, \I_{2T} \quad \0 \, \big] \\[0.5mm] 
 \boldsymbol{\upsilon}_{j,2} = \Null{\big[ 
  \mathbf{\tilde{H}}_{j,j-1} \big( :,1 \big), 
  \mathbf{\tilde{H}}_{j,j+1} \big( :,2:3 \big)
\big]} \\[2.5mm]
 \boldsymbol{\upsilon}_{j,3} = \Null{\big[ 
  \mathbf{\tilde{H}}_{j,j-1} \big( :,1:2 \big), 
  \mathbf{\tilde{H}}_{j,j+1} \big( :,3 \big)
\big]} \\[0.5mm]
 \boldsymbol{\upsilon}_{j,4} = \big[ \, \0 \quad \I_{2T} \, \big],
 \end{matrix}
\end{IEEEeqnarray}
where $\mathbf{A}(:,b:c)$ gives the matrix resulting from picking the entries of $\An$ from block column $b$ to $c$, and $\I_{2T} \Rmat{2T}{2T}$, $\0 \Rmat{2T}{2Tp}$ are the identity and all-zero matrices.

\def\arraystretch{1}
\setlength{\arraycolsep}{0.5pt}
\begin{figure*}
\setcounter{equation}{39}
\def\arraystretch{1.5}
\setlength{\arraycolsep}{0.5pt}
\begin{IEEEeqnarray}{l}
\left[\mathbf{\tilde{H}}_{j,j-1},\mathbf{\tilde{H}}_{j,j+1}\right] = 
\begin{bmatrix}
	\Cn \big( \tilde{h}_{j,j-1}^{1,1} \big) & \0 	& \0
	& \0 & \0 & \0
	\\
	  \Cn \big( \tilde{h}_{j,j-1}^{2,1} \big) 
   & \Cn \big( \tilde{h}_{j,j-1}^{2,2} \big) 
	& \Cn \big( \tilde{h}_{j,j-1}^{2,3} \big)
	& \Cn \big( \tilde{h}_{j,j+1}^{2,1} \big)
	& \0
 	& \Cn \big( \tilde{h}_{j,j+1}^{2,3} \big)
	\\
	  \Cn \big( \tilde{h}_{j,j-1}^{3,1} \big) 
	& \0 
	& \Cn \big( \tilde{h}_{j,j-1}^{3,3} \big)
	& \Cn \big( \tilde{h}_{j,j+1}^{3,1} \big) 
   & \Cn \big( \tilde{h}_{j,j+1}^{3,2} \big) 
	& \Cn \big( \tilde{h}_{j,j+1}^{3,3} \big)
	\\
	\0 & \0 & \0
	& \0
	& \0 
	& \Cn \big( \tilde{h}_{j,j+1}^{4,3} \big)
\end{bmatrix}
\label{eq:estructuraPreFinal_H_p=3}
\end{IEEEeqnarray}
\begin{IEEEeqnarray}{l}
\boldsymbol{\Upsilon}_j \left[\mathbf{\tilde{H}}_{j,j-1},\mathbf{\tilde{H}}_{j,j+1}\right] \!= \! 
\begin{bmatrix}
	\Cn \big( {h}_{j,j-1}^{1,1} \big) & \0 	& \0
	& \0 & \0 & \0
	\\
	  \0
   & \Cn \big( {h}_{j,j-1}^{2,2} \big) 
	& \Cn \big( {h}_{j,j-1}^{2,3} \big)
	& \Cn \big( {h}_{j,j+1}^{2,1} \big)
	& \0
 	& \0
	\\
	  \0 
	& \0 
	& \Cn \big( {h}_{j,j-1}^{3,3} \big)
	& \Cn \big( {h}_{j,j+1}^{3,1} \big) 
   & \Cn \big( {h}_{j,j+1}^{3,2} \big) 
	& \0
	\\
	\0 & \0 & \0
	& \0
	& \0 
	& \Cn \big( {h}_{j,j+1}^{4,3} \big)
\end{bmatrix}
\label{eq:estructuraFinal_H_p=3}
\end{IEEEeqnarray}
% \vspace{-5mm}
\hrule
\vspace{-5mm}
\end{figure*}

%%%%%%%%%%%%%%%%%%%%%%%%%%%%%%%%%%%%%%%%%%%%%%%%%%%%%%%%%
%%%%%%%%%%%%%%%%%%%%%%%%%%%%%%%%%%%%%%%%%%%%%%%%%%%%%%%%%
\section{Proof of Lemma 1}
\label{sec:appendixB}
We will prove the lemma for the system of equations defined for $j=1$. Cases $j=2,3$ can be similarly handled, due to symmetry of the problem. Therefore, we drop the supraindex $j$ and write ${\boldsymbol{\lambda}_i, i=1,\ldots5}$ to simplify notation. Some rank-preserving transformations are applied to $\mathbf{G}_j$, such that (\ref{eq:29Glanda}) for $j=1$ can be written as follows:
\setcounter{equation}{42}
\begin{IEEEeqnarray}{c}
{ \Ch{1,1}{1,1} {\An_2}{{\boldsymbol{\lambda }}_1} + {\An_1}{{\boldsymbol{\lambda }}_3} = {\0}}  \nonumber,
\\[-2mm]
{\Ch{1,1}{2,1} {\An_2}{{\boldsymbol{\lambda }}_1} + 
\Ch{1,1}{2,2}{\An_1}{{\boldsymbol{\lambda }}_2} + 
{\An_3}{{\boldsymbol{\lambda }}_4} = {\0}}, \qquad \label{eq:41}
\\[-2mm] \nonumber
{\Ch{1,1}{3,2}{\An_1}{{\boldsymbol{\lambda }}_2} + 
{\An_2}{{\boldsymbol{\lambda }}_5} = {\0}},
\end{IEEEeqnarray}
which can be simplified by introducing (\ref{eq:30Adesign}), and by means of linear independence among $\mathbf{b}_i$. 
For instance, consider all equations corresponding to $\mathbf{B}_{1:2}$ in (\ref{eq:41}):
\begin{IEEEeqnarray}{c}
\label{eq:4,3_eq1}
\Ch{1,1}{1,1} \mathbf{b}_q \, \lambda _1 
\left( q \right) + \mathbf{b}_q \, \lambda _3
\left( q \right) = \0, \,  \\[-2mm]
\label{eq:4,3_eq2}
\Ch{1,1}{2,1} \mathbf{b}_q \, \lambda _1 
\left( q \right) + 
\Ch{1,1}{2,2} \mathbf{b}_q \, \lambda _2 
\left( q \right) = \0, \,  \\[-2mm]
\label{eq:4,3_eq3}
\Ch{1,1}{3,2} \mathbf{b}_q \, \lambda _2 
\left( q \right) +
\mathbf{b}_q \, \lambda _5
\left( q \right) = \0, \,
\end{IEEEeqnarray}
with $q=1,2$. Each of such equations can be simplified as follows. Let us define: 
\begin{IEEEeqnarray}{c}
\mathbf{\tilde b}_q = 
\begin{bmatrix}
b_q \left( 1 \right) + j b_q \left( 2 \right) \quad & b_q \left( 3 \right) + j b_q \left( 4 \right) \quad& \ldots \quad & b_q \left( 9 \right) + j b_q \left( 10 \right)
\end{bmatrix}^T ,
\label{eq:44}
\end{IEEEeqnarray}
where $\mathbf{b}_q = 
\begin{bmatrix}
b_q \left( 1 \right) \quad & b_q \left( 2 \right) \quad & \ldots \quad& b_q \left( 10 \right)
\end{bmatrix}^T$, $j=\sqrt{-1}$ stands for the imaginary unit, and $q=1,2$. Then, as in \cite{ACS}, we can write (\ref{eq:4,3_eq1})-(\ref{eq:4,3_eq3}) in terms of $\mathbf{\tilde b}_q$. For instance, (\ref{eq:4,3_eq1}) can be rewritten as follows:
\begin{IEEEeqnarray}{c}
\begin{matrix}
\left| h_{1,1}^{1,1} \right|{e^{j \phi _{1,1}^{1,1}}}{{{\mathbf{\tilde b}}}_q} \,{\lambda _1}\left( q \right) + {{{\mathbf{\tilde b}}}_q}\,{\lambda _3}\left( q \right) = {\0}
\quad \Rightarrow \quad
\left| h_{1,1}^{1,1} \right|{e^{j \phi _{1,1}^{1,1}}}{\lambda _1}\left( q \right) + {\lambda _3}\left( q \right) = 0
\end{matrix},
\label{eq:45}
\end{IEEEeqnarray}
with $q = 1,2$. Hence, equating real and imaginary parts of each equation to zero, we have:
\begin{IEEEeqnarray}{c}
\begin{matrix}
\left| h_{1,1}^{1,1} \right|\sin \left( {\phi _{1,1}^{1,1}} \right)\lambda _1(q) = 0,
\\
\left| h_{1,1}^{1,1} \right|\cos \left( {\phi _{1,1}^{1,1}} \right)\lambda _1(q) + \lambda _3(q) = 0,
\end{matrix}
\label{eq:46}
\end{IEEEeqnarray}
with $q = 1,2$. The set containing all the possible values such that $\left| h_{1,1}^{1,1} \right|\sin \left( {\phi _{1,1}^{1,1}} \right) = 0$ is a countable set, thus it has zero measure \cite{MeasureProbability}. By randomness arguments the only solution is $\lambda _r(q) = 0, r = 1,3, q = 1,2$. Applying this methodology to all equations derived from all groups of columns of $\mathbf{B}$, one finds out that all rank multipliers must be zero.

We present an alternative way to see that the rank multipliers associated to $\mathbf{B}_{1:2}$  must be zero. Instead of developing (\ref{eq:4,3_eq1}) only, let us write all equations (\ref{eq:4,3_eq1})-(\ref{eq:4,3_eq3}) in the form of (\ref{eq:46}). Then, equating imaginary parts to zero, some rank multipliers can be determined as the solution to
\begin{IEEEeqnarray}{c}
\begin{bmatrix}
 \,\left| h_{1,1}^{1,1} \right|\sin \left( {\phi _{1,1}^{1,1}} \right) 
& 0\,\\
 \,\left| h_{1,1}^{2,1} \right|\sin \left( {\phi _{1,1}^{2,1}} \right) 
 & \left| h_{1,1}^{2,2} \right|\sin \left( {\phi _{1,1}^{2,2}} \right) 
 \,\\
 \, 0 & \left| h_{1,1}^{3,2} \right|\sin \left( {\phi _{1,1}^{3,2}} \right) 
 \,\\
\end{bmatrix} \!\!
\begin{bmatrix}
 \,\lambda _1(q) \,\\
 \,\lambda _2(q) \,\\
\end{bmatrix}
= \0.
\label{eq:eliminationMatrix}
\end{IEEEeqnarray}
We will refer to the $3 \times 2$ matrix at the left-hand side of (\ref{eq:eliminationMatrix}) as an \textit{elimination matrix}. As long as we can ensure it has no right null space, all rank multipliers in (\ref{eq:eliminationMatrix}) can be set to zero. In this case, this is trivially ensured by means of randomness arguments. Likewise, using the real counterpart of (\ref{eq:eliminationMatrix}), we have ${\lambda _i(q)=0, i=3,5},$ $q=1,2$. By the same rationale applied to each group of columns of $\mathbf{B}$, we obtain an elimination matrix for each case, and it is easy to check that none of them has right null space, thus all rank multipliers are definitely equal to zero.

So far we have proved that considering ACS is sufficient for achieving a full rank SSM. In what follows, we explain why it is necessary when using the scheme based on alignment chains. In this regard, notice that if only symbol extensions in time are employed, a set of equations similar to (\ref{eq:4,3_eq1})-(\ref{eq:4,3_eq3}) is obtained, and we have
\begin{IEEEeqnarray}{c}
\def\arraystretch{1}
\setlength{\arraycolsep}{7pt}
\begin{matrix}
h_{1,1}^{1,1} \lambda _1(q) + \lambda _3(q) = 0 \\
h_{1,1}^{2,1}  \lambda _1 
\left( q \right) + 
h_{1,1}^{2,2}  \lambda _2 
\left( q \right) = 0   \\
h_{1,1}^{3,2}  \lambda _2 
\left( q \right) + \lambda _5(q) = 0 
\end{matrix} \quad \Rightarrow \quad 
\begin{bmatrix}
 h_{1,1}^{1,1} & 0 & 1 & 0 \\
 h_{1,1}^{2,1} & h_{1,1}^{2,2} & 0 & 0 \\
 0 & h_{1,1}^{3,2} & 0 & 1 
\end{bmatrix}
\begin{bmatrix}
  \lambda_1(q) \\
 \lambda_2(q) \\
 \lambda_3(q) \\
 \lambda_5(q) 
\end{bmatrix}
= \0,
\label{eq:eliminationMatrixOnlyT}
\end{IEEEeqnarray}
\\
with $q=1,2$, where all $\mathbf{C}(\cdot)$ disappear since channel elements are written in the extended model as scaled identity matrices, and the rank multipliers are now complex magnitudes. In this case, the elimination matrix is a $3 \times 4$ full-row rank matrix, thus there exists at least one non-zero solution. Consequently, the SSM becomes rank deficient since there are some rank multipliers different from zero and thus desired signals cannot be separated from interference.

%%%%%%%%%%%%%%%%%%%%%%%%%%%%%%%%%%%%%%%%%%%%%%%%%%%%%%%%%
%%%%%%%%%%%%%%%%%%%%%%%%%%%%%%%%%%%%%%%%%%%%%%%%%%%%%%%%%

\section{Proof of Lemma 2}
\label{sec:appendixB1}
Due to similarity with the proof for $p=2$, we elaborate a sketch of the proof for $p=3$ and provide intuition of the proof for $p=4,5,6$ by means of examples of its elimination matrices.

The SSM for $p=3$ is constructed by using 
(\ref{eq:prodCreuatsa})-(\ref{eq:prodCreuatsb}). As before, without loss of generality we consider receiver 1 only.
In this case, after applying some full-rank linear transformations to the SSM, the following system of four equations is obtained:
 \begin{IEEEeqnarray*}{l}
 \small
\left[ \Ch{1,1}{1,1} - 
  \C{ \alpha^\text{des}_{1,1} }
\right] 
\An_3 \boldsymbol{\lambda }_1  +
\Ch{1,1}{1,2} \An_2 \boldsymbol{\lambda}_2   
+\left[ \Ch{1,1}{1,3} - 
  \C{ \alpha^\text{des}_{1,2} }
\right] 
{\An_1}{{\boldsymbol{\lambda }}_3}  +
\Ch{1,3}{1,1} {\An_2}{\boldsymbol{\lambda }}_4 = {\0},
 \\
 \small
\left[ \Ch{1,1}{2,1} - 
  \C{ \alpha^\text{des}_{2,1} }
\right] 
{\An_3}{{\boldsymbol{\lambda}_1}} +  
\Ch{1,1}{2,2} \An_2 \boldsymbol{\lambda}_2 
-\C{ \alpha^\text{int}_{1} }
{\An_2}{\boldsymbol{\lambda }}_4 +
\An_1{{\boldsymbol{\lambda }}_5} = {\0},
\end{IEEEeqnarray*}
 \begin{IEEEeqnarray*}{l}
 \small
\Ch{1,1}{3,2} \An_2 \boldsymbol{\lambda}_2 
+ \left  [ \Ch{1,1}{3,3} - 
  \C{ \alpha^\text{des}_{3,2} }
\right]  
{\An_1}{{\boldsymbol{\lambda }}_3} 
 +
\An_3{{\boldsymbol{\lambda }}_6} 
-\C{ \alpha^\text{int}_{2} }
{\An_2}{\boldsymbol{\lambda }}_7= {\0},
\\
\small
\left[ \Ch{1,1}{41} - 
  \C{ \alpha^\text{des}_{4,1} }
\right]  
{\An_3}{{\boldsymbol{\lambda }}_1} +
\Ch{1,1}{42} \An_2 \boldsymbol{\lambda}_2  
+{\left[ \Ch{1,1}{4,3} - 
  \C{ \alpha^\text{des}_{4,2} }
\right] 
{\An_1}{{\boldsymbol{\lambda }}_3}}
+ 
\Ch{1,2}{4,3} \An_2 \boldsymbol{\lambda}_7  = {\0},
 \label{eq:41_p=3}
\end{IEEEeqnarray*}
where the SPBs are chosen as in (\ref{eq:37}), i.e:
\begin{IEEEeqnarray}{c}
\setlength{\arraycolsep}{3pt}
\An_1 = \left[
\begin{matrix}
	\mathbf{B}_{1:2} & \mathbf{B}_{3:6} & \mathbf{B}_{7:8}
\end{matrix}
\right],  \quad 
\An_2 =
\begin{bmatrix}
	\mathbf{B}_{1:2} & \mathbf{B}_{9:12} & \mathbf{B}_{13:14}
\end{bmatrix},
\quad
\An_3 =
\begin{bmatrix}
	\mathbf{B}_{3:6} & \mathbf{B}_{9:12}
\end{bmatrix},
\label{eq:A_p3}
\end{IEEEeqnarray}
% \vspace{-2mm}
% \\
\vspace{-3mm}
and
\vspace{-3mm}
\begin{IEEEeqnarray*}{c c c c c c c}
 \alpha^\text{des}_{q,1} = \frac{\hel{1,1}{q,2}\hel{3,1}{3,1}}{\hel{3,1}{3,2}} ,  
 & \quad &
 \alpha^\text{des}_{q,2} = \frac{\hel{1,1}{q,2}\hel{2,1}{2,3}}{\hel{2,1}{2,2}} ,  
 & \quad &
 \alpha^\text{int}_{1} = \frac{\hel{1,3}{2,2}\hel{2,3}{3,1}}{\hel{2,3}{3,2}} , 
 & \quad &
 \alpha^\text{int}_{2} = \frac{\hel{1,2}{3,2}\hel{3,1}{2,3}}{\hel{3,1}{2,2}}. 
\end{IEEEeqnarray*}

The SSM is full-rank as long as all rank multipliers ${\boldsymbol{\lambda}_i, i=1,\ldots,7}$ are equal to zero. For instance, consider the elimination matrix in (\ref{eq:eliminationMatrixp3}), obtained for the group $Z$ (see (\ref{eq:sets})) after applying similar steps as in Appendix \ref{sec:appendixB}, and equating imaginary parts to zero.
Notice that this elimination matrix is full rank almost surely, since each row contains at least one element of the direct channel. Therefore, all rank multipliers involved in (\ref{eq:eliminationMatrixp3}) can be set to zero.

Similar ideas apply to cases $p=4,5,6$. For the sake of brevity, we show only the elimination matrix analogous to (\ref{eq:eliminationMatrixp3}) for each of those cases at the next page, where following similar arguments discussed above, it can be ensured that all the elimination matrices are full rank, they have no right null space, and thus all involved rank multipliers can be set to zero. To simplify notation we have used the function $\psi(a,b)$, defined as the sum of the sinusoidal functions corresponding to the position $(a,b)$ of each elimination matrix.

\begin{figure*}
 \begin{IEEEeqnarray}{c}
 \small
\begin{bmatrix}
 \left| h_{1,1}^{1,2} \right|\sin \left( {\phi _{1,1}^{1,2}} \right) 
 &
 \left| h_{1,1}^{1,3} \right|\sin \left( {\phi _{1,1}^{1,3}} \right) -
 \left|\alpha^\text{des}_{1,2} \right|\sin \left( {\alpha^\text{des}_{1,2}} \right)
 & \left| h_{1,3}^{1,1} \right|\sin \left( {\phi _{1,3}^{1,1}} \right)
 & 0\\
 \left| h_{1,1}^{2,2} \right|\sin \left( {\phi _{1,1}^{2,2}} \right) & 0
 & -\left|\alpha^\text{int}_{1} \right|\sin \left( \alpha^\text{int}_{1} \right) 
 & 0 \\
  \left| h_{1,1}^{3,2} \right|\sin \left( {\phi _{1,1}^{3,2}} \right) & 
  \left| h_{1,1}^{3,3} \right|\sin \left( {\phi _{1,1}^{3,3}} \right) -
 \left|\alpha^\text{des}_{3,2} \right|\sin \left( {\alpha^\text{des}_{3,2}} \right)
 & -\left|\alpha^\text{int}_{1} \right|\sin \left( \alpha^\text{int}_{1} \right) 
  & -\left|\alpha^\text{int}_{2} \right|\sin \left( \alpha^\text{int}_{2} \right)
 \\
\left| h_{1,1}^{42} \right|\sin \left( {\phi _{1,1}^{42}} \right) 
 &
 \left| h_{1,1}^{4,3} \right|\sin \left( {\phi _{1,1}^{4,3}} \right) -
 \left|\alpha^\text{des}_{4,2} \right|\sin \left( {\alpha^\text{des}_{4,2}} \right)
 & \left| h_{1,2}^{4,3} \right|\sin \left( {\phi _{1,2}^{4,3}} \right)
 & 0\\
\end{bmatrix} \!\!\! \!
\begin{bmatrix}
 \lambda _2(q) \\
 \lambda _3(q) \\
 \lambda _4(q) \\
 \lambda _7(q) 
\end{bmatrix}
\!= \! \0 
\nonumber \vspace{-3mm} \\  \vspace{-1mm}
\label{eq:eliminationMatrixp3}
\end{IEEEeqnarray}
\hrule

\begin{IEEEeqnarray*}{l}
p=4 :\\[3mm]
 \begin{bmatrix}
 0 & \elim{1}{2} & \elim{1}{3} & 0 
 \\
 \elim{2}{1} & 0 &  \elim{2}{3} &  0
 \\
 0 & \elim{3}{2} &  0 &  \elim{3}{4}
 \\
 \elim{4}{1} & \elim{4}{2} &  \elim{4}{3} &  \elim{4}{4}
 \\
 \elim{5}{1} & 0 &  \elim{5}{3} & \elim{5}{4}
\end{bmatrix} 
\begin{bmatrix}
 \lambda _1(q) \\
 \lambda _3(q) \\
 \lambda _4(q) \\
 \lambda _9(q) 
\end{bmatrix}
=\0
 \\[4mm]
 p=5 :\\[3mm]
 \begin{bmatrix}
 \elim{1}{1} & 0 &  0 &  0 & \elim{1}{5} & 0
 \\
 \elim{2}{1} & \elim{2}{2} &  \elim{2}{3} &  \elim{2}{4} & \elim{2}{5} & 0
 \\
 0 & \elim{3}{2} &  0 &  \elim{3}{4} & \elim{3}{5} &  0
 \\
 \elim{4}{1} & 0 & \elim{4}{3} &  0 &  0 & \elim{4}{6}
 \\
 \elim{5}{1} & \elim{5}{2} &  \elim{5}{3} & \elim{5}{4} & 0 & \elim{5}{6}
 \\
 0 & \elim{6}{2} & 0 & 0 & 0 & \elim{6}{6}
\end{bmatrix}
\begin{bmatrix}
 \lambda _1(q) \\
 \lambda _2(q) \\
 \lambda _4(q) \\
 \lambda _5(q) \\
 \lambda _9(q) \\
 \lambda _{11}(q)
\end{bmatrix}
=\0
 \\[4mm]
 p=6 :\\[3mm]
\begin{bmatrix}
 0 & 0 &  0 &  \elim{1}{4} & \elim{1}{5} & 0 & 0
 \\
 0 & 0 &  \elim{2}{3} &  0 & \elim{2}{5} & \elim{2}{6} & 0
 \\
 \elim{3}{1} & \elim{3}{2} &  \elim{3}{3} &  \elim{3}{4} & \elim{3}{5} &  \elim{3}{6} & 0 
 \\
 0 & \elim{4}{2} & 0 &  \elim{4}{4} &  0 & \elim{4}{6} & 0
 \\
 \elim{5}{1} & 0 &  \elim{5}{3} & 0 & 0 & 0 & \elim{5}{7}
 \\
 \elim{6}{1} & \elim{6}{2} & \elim{6}{3} & \elim{6}{4} & 0 & 0 & \elim{6}{7}
 \\
 0 & \elim{7}{2} & 0 & \elim{7}{4} & 0 & 0 & \elim{7}{7}
\end{bmatrix}
\begin{bmatrix}
 \lambda _2(q) \\
 \lambda _3(q) \\
 \lambda _5(q) \\
 \lambda _6(q) \\
 \lambda _7(q) \\
 \lambda _{8}(q) \\
\lambda _{13}(q) 
\end{bmatrix}
=\0
\end{IEEEeqnarray*}
\\[10mm]
\end{figure*}
\newpage 
\bibliographystyle{IEEEtranTCOM}
\bibliography{../../../papers/_referenciesMarc}

% Generated by IEEEtranTCOM.bst, version: 1.13 (2008/09/30)
\begin{thebibliography}{10}
\baselineskip 12pt
\providecommand{\url}[1]{#1}
\csname url@samestyle\endcsname
\providecommand{\newblock}{\relax}
\providecommand{\bibinfo}[2]{#2}
\providecommand{\BIBentrySTDinterwordspacing}{\spaceskip=0pt\relax}
\providecommand{\BIBentryALTinterwordstretchfactor}{4}
\providecommand{\BIBentryALTinterwordspacing}{\spaceskip=\fontdimen2\font plus
\BIBentryALTinterwordstretchfactor\fontdimen3\font minus
  \fontdimen4\font\relax}
\providecommand{\BIBforeignlanguage}[2]{{%
\expandafter\ifx\csname l@#1\endcsname\relax
\typeout{** WARNING: IEEEtran.bst: No hyphenation pattern has been}%
\typeout{** loaded for the language `#1'. Using the pattern for}%
\typeout{** the default language instead.}%
\else
\language=\csname l@#1\endcsname
\fi
#2}}
\providecommand{\BIBdecl}{\relax}
\BIBdecl

\bibitem{AlignmentChainsTrans}
C.~Wang, T.~Gou, and S.~Jafar, ``{Subspace Alignment Chains and the Degrees of
  Freedom of the Three-User MIMO Interference Channel},'' \emph{IEEE Trans.
  Inf. Theory}, vol.~60, pp. 2432 -- 2479, May 2014.

\bibitem{Birk&Kol}
Y.~Birk and T.~Kol, ``{Informed-source coding-on-demand (ISCOD) over broadcast
  channels},'' in \emph{IEEE INFOCOM}, Mar. 1998.

\bibitem{Maddah-Ali2008}
M.~Maddah-Ali, A.~Motahari, and A.~Khandani, ``{Communication Over MIMO X
  Channels: Interference Alignment, Decomposition, and Performance Analysis},''
  \emph{IEEE Trans. Inf. Theory}, Aug. 2008.

\bibitem{CJ}
V.~Cadambe and S.~Jafar, ``{Interference Alignment and Degrees of Freedom of
  the $K$-User Interference Channel},'' \emph{IEEE Trans. Inf. Theory},
  vol.~54, pp. 3425 -- 3441, Aug. 2008.

\bibitem{Yetis2010}
C.~M. Yetis, T.~Gou, S.~Jafar, and A.~H. Kayran, ``{On feasibility of
  interference alignment in MIMO interference networks},'' \emph{IEEE Trans.
  Signal Process.}, Sep. 2010.

\bibitem{Razaviyayn2012}
M.~Razaviyayn, G.~Lyubeznik, and Z.-Q. Luo, ``{On the Degrees of Freedom
  Achievable Through Interference Alignment in a MIMO Interference Channel},''
  \emph{IEEE Trans. Signal Process.}, Feb. 2012.

\bibitem{Gou2010_DoF_MN}
T.~Gou and S.~Jafar, ``{Degrees of Freedom of the $K$-User $M \times N$ MIMO
  Interference Channel},'' \emph{IEEE Trans. Inf. Theory}, Dec. 2010.

\bibitem{GenieChainsArxiv}
C.~{Wang}, H.~{Sun}, and S.~{Jafar}, ``{Genie Chains: Exploring Outer Bounds on
  the Degrees of Freedom of MIMO Interference Networks},'' \emph{ArXiv
  e-prints}, vol. arXiv:1404.2258v1 [cs.IT], Apr. 2014.

\bibitem{Reciprocity}
K.~Gomadam, V.~Cadambe, and S.~Jafar, ``{Approaching the Capacity of Wireless
  Networks through Distributed Interference Alignment},'' in \emph{IEEE
  GLOBECOM}, Nov. 2008.

\bibitem{ACS}
V.~Cadambe, S.~Jafar, and C.~Wang, ``{Interference Alignment With Asymmetric
  Complex Signaling - Settling the Host-Madsen-Nosratinia Conjecture},''
  \emph{IEEE Trans. Inf. Theory}, Sept. 2010.

\bibitem{Geometry2011}
G.~Bresler, D.~Cartwright, and D.~Tse, ``{Geometry of the 3-user MIMO
  interference channel},'' in \emph{Allerton}, Sep. 2011.

\bibitem{LayeredIA}
S.~Mahboubi, A.~Motahari, and A.~Khandani, ``{Layered Interference Alignment:
  Achieving the total DoF of MIMO X-channels},'' in \emph{IEEE ISIT}, Jun.
  2010.

\bibitem{Ghasemi2010}
A.~Ghasemi, A.~Motahari, and A.~Khandani, ``{Interference alignment for the
  $K$-user MIMO interference channel},'' in \emph{IEEE ISIT}, Jun. 2010.

\bibitem{Ordentlich2013}
O.~Ordentlich and U.~Erez, ``{On the Robustness of Lattice Interference
  Alignment},'' \emph{IEEE Trans. Inf. Theory}, May 2013.

\bibitem{ErgodicIA}
B.~Nazer, M.~Gastpar, S.~Jafar, and S.~Vishwanath, ``{Ergodic Interference
  Alignment},'' \emph{IEEE Trans. Inf. Theory}, Oct. 2012.

\bibitem{OIA}
J.~H. Lee and W.~Choi, ``{On the Achievable DoF and User Scaling Law of
  Opportunistic Interference Alignment in 3-Transmitter MIMO Interference
  Channels},'' \emph{IEEE Trans. Wireless Commun.}, Jun. 2013.

\bibitem{ACS_4users}
C.~Lameiro and I.~Santamaria, ``{Degrees-of-freedom for the 4-user SISO
  interference channel with improper signaling},'' in \emph{IEEE ICC}, Jun.
  2013.

\bibitem{Zeng_defRank_MIMO3}
Y.~{Zeng}, X.~{Xu}, Y.~L. {Guan}, and E.~{Gunawan}, ``{On the Achievable
  Degrees of Freedom for the 3-User Rank-Deficient MIMO Interference
  Channel},'' \emph{IEEE Trans. Wireless Commun.}, Apr. 2014.

\bibitem{AlignmentChains_ISIT}
C.~Wang, T.~Gou, and S.~Jafar, ``{Subspace alignment chains and the degrees of
  freedom of the three-user MIMO interference channel},'' in \emph{IEEE ISIT},
  Jul. 2012.

\bibitem{ACSnoPublicat}
------, ``{On Optimality of Linear Interference Alignment for the Three-User
  MIMO Interference Channel with Constant Channel Coefficients},'' in
  \emph{eScholarship Univ. of California}, Oct. 2011, available at:
  http://escholarship.org/uc/item/6t14c361.

\bibitem{MeasureProbability}
M.~Capinski and P.~Kopp, \emph{{Measure, integral and probability}},
  2003rd~ed., ser. Springer undergraduate mathematics Series.\hskip 1em plus
  0.5em minus 0.4em\relax Londres: Springer, 1999.

\end{thebibliography}

\end{document}